\documentclass[12pt]{article}

\usepackage[utf8]{inputenc}
\usepackage{amsmath,amssymb}
\usepackage{slashed}
\usepackage{authblk}
\usepackage{xcolor}
\usepackage{hyperref}
\usepackage{cite}

\date{\today}

\newcommand{\swb}{\bar{s}^2_\smallw}
\newcommand{\cwb}{\bar{c}^2_\smallw}
\newcommand{\mw}{M_{\rm W}} 
\newcommand{\mz}{M_{\rm Z}} 
\newcommand{\mh}{M_{\rm H}}
\newcommand{\mt}{m_t}
\newcommand{\xt}{X_t}
\newcommand{\sss}[1]{\scriptscriptstyle{#1}}

\newcommand{\smallw}{{\scriptscriptstyle W}}
\newcommand{\SANC}{\texttt{SANC}}

\newcommand{\MCSANCee}{\texttt{MCSANCee}}
\newcommand{\ReneSANCe}{\texttt{ReneSANCe}}
\newcommand{\WHIZARD}{\texttt{WHIZARD}}
\newcommand{\CalcHEP}{\texttt{CalcHEP}}

\newcommand{\tffgg} {{\tilde F}_{\sss \gamma}}
\newcommand{\tffld}{{\tilde F}_{\sss LD}}
\newcommand{\tffqd}{{\tilde F}_{\sss QD}}
\newcommand{\tffqq}{{\tilde F}_{\sss QQ}}
\newcommand{\tffll}{{\tilde F}_{\sss LL}}
\newcommand{\tfflq}{{\tilde F}_{\sss LQ}}
\newcommand{\tffql}{{\tilde F}_{\sss QL}}
\newcommand{\ffgg} {{F}_{\sss \gamma}}
\newcommand{\ffld}{{F}_{\sss LD}}
\newcommand{\ffqd}{{F}_{\sss QD}}
\newcommand{\ffqq}{{F}_{\sss QQ}}
\newcommand{\ffll}{{F}_{\sss LL}}
\newcommand{\fflq}{{F}_{\sss LQ}}
\newcommand{\ffql}{{F}_{\sss QL}}

\newcommand{\mel}{m_e}
\newcommand{\rhoone}{\rho^{(1)}}
\newcommand{\rhotwo}{\rho^{(2)}}
\newcommand{\mmo}{m_\mu}
\newcommand{\qel}{Q_e}
\newcommand{\qmo}{Q_\mu}
\newcommand{\cpl}{c^+}
\newcommand{\cmi}{c^-}

\newcommand{\vmael}{\delta_{e}}
\newcommand{\vmamo}{\delta_{\mu}}
\newcommand{\ael}{I^{(3)}_e}
\newcommand{\amo}{I^{(3)}_\mu}
\newcommand{\chizt}{\chi_{\sss{Z}}(t)}
\newcommand{\nll}{\nonumber\\}
\newcommand{\Litwo}{\mbox{${\rm{Li}}_{2}$}}
\newcommand{\bqa}{\begin{eqnarray}}
\newcommand{\eqa}{\end{eqnarray}}
\newcommand{\ds }{\displaystyle}
\newcommand{\xel }{x_e}
\newcommand{\xmo }{x_\mu}
\newcommand{\relms}{m_e^2}
\newcommand{\rmoms}{m_\mu^2}
\newcommand{\lnlaom}{\ln\left(\frac{ 4 \omega^2}{\lambda^2}\right)}
\newcommand{\ts }{t}
\newcommand{\us }{u}
\newcommand{\rms }{r_{s}}
\newcommand{\rmt }{r_{t}}
\newcommand{\sqs }{\sqrt{s}}
\newcommand{\sw}{s^2_\smallw}
\newcommand{\cw}{c^2_\smallw}

\newcommand{\stw}{s_\smallw}
\newcommand{\ctw}{c_\smallw}

\usepackage{scalerel}

\newcommand{\rbrket}{\hstretch{1.43}{\rbrack}}
\newcommand{\lbrket}{\hstretch{1.43}{\lbrack}}

\newcommand{\SpIA}[1]{\stretchrel*{\lvert }{#1|}#1\stretchrel*{\rangle}{#1|}}
\newcommand{\SpIB}[1]{\stretchrel*{\lvert }{#1|}#1\stretchrel*{\rbrket}{#1|}}
\newcommand{\SpAI}[1]{\stretchrel*{\langle}{#1|}#1\stretchrel*{\rvert }{#1|}}
\newcommand{\SpBI}[1]{\stretchrel*{\lbrket}{#1|}#1\stretchrel*{\rvert }{#1|}}

\newcommand{\SpIAoAI}[2]{\stretchrel*{\lvert}{#1|#2}#1\stretchrel*[550]{\rangle}{#1|#2}\!\stretchrel*{\langle}{#1|#2}#2\stretchrel*{\rvert}{#1|#2}}

\newcommand{\SpIBoBI}[2]{\stretchrel*{\lvert}{#1|#2}#1\stretchrel*[550]{\rbrket}{#1|#2}\!\stretchrel*{\lbrket}{#1|#2}#2\stretchrel*{\rvert}{#1|#2}}

\newcommand{\SpAIA}[2]{\stretchrel*[550]{\langle}{#1|#2}#1\stretchrel*{\vert}{#1|#2}#2\stretchrel*{\rangle}{#1|#2}}
\newcommand{\SpBIB}[2]{\stretchrel*[550]{\lbrket}{#1|#2}#1\stretchrel*{\vert}{#1|#2}#2\stretchrel*{\rbrket}{#1|#2}}

\newcommand{\SpBIIA}[3]{\stretchrel*[550]{\lbrket}{#1|#2#3}#1\stretchrel*{\vert}{#1|#2#3}#2\stretchrel*{\vert}{#1|#2#3}#3\stretchrel*{\rangle}{#1|#2#3}}
\newcommand{\SpAIIB}[3]{\stretchrel*[550]{\langle}{#1|#2#3}#1\stretchrel*{\vert}{#1|#2#3}#2\stretchrel*{\vert}{#1|#2#3}#3\stretchrel*{\rbrket}{#1|#2#3}}

\DeclareMathOperator{\Tr}{Tr}

\newcommand{\hel}[5]{{}_{#1}{}_{#2}{}_{#3}{}_{#4}{}_{#5}}
\newcommand{\labhel}[6]{{}^{#1}_{{#2}{#3}{#4}{#5}{#6}} }

\usepackage{xcolor}

\begin{document}

\title{Electroweak effects in polarized muon-electron scattering}

\author[1,2]{A.B.~Arbuzov}

\author[1]{S.G.~Bondarenko}

\author[3]{L.V.~Kalinovskaya}

\author[3]{L.A.~Rumyantsev}

\author[3,4]{V.L.~Yermolchyk}

\affil[1]{\small Bogoliubov Laboratory of Theoretical Physics, Joint Institute for Nuclear Research, Dubna, 141980 Russia}

\affil[2]{\small Department of higher mathematics, Dubna State University, Dubna, 141982 Russia}

\affil[3]{\small Dzhelepov Laboratory of Nuclear Problems, Joint Institute for Nuclear Research, Dubna, 141980 Russia}

\affil[4]{\small Institute for Nuclear Problems, Belarusian State University,
Minsk, 220006  Belarus}

\maketitle

\begin{abstract}
A theoretical description of elastic polarized muon-electron scattering is presented.
Complete one-loop electroweak radiative corrections are calculated 
with taking into account the exact dependence on the muon mass.
The effects due to some higher-order corrections and the electroweak scheme dependence 
are analyzed.
The case of longitudinally polarized fermions in the initial state is investigated.
Analytical results are derived with the help of the \SANC~system.
Numerical results are presented for unpolarized and polarized cross sections.
Calculations are realized in the Monte Carlo integrator \MCSANCee~ and generator \ReneSANCe~
which allow the implementation of any experimental cuts used in the analysis of elastic
$\mu - e$ scattering data.
\end{abstract}

\section{Introduction \label{Sect:Intro}}

Muon-electron scattering is one of the most pure QED processes
of elementary particle interactions. So its cross section can be 
calculated perturbatively within the Standard Model with a very high precision.
On the other hand, this process is suited for high-precision 
experimental measurements since it has a clear detector signature.
This allows to use this process for studies of electroweak (EW) 
and strong interaction effects, given the QED part is fully understood.

Radiative corrections in polarized electron muon elastic scattering
were considered by \cite{Kukhto:1987uj}
and later fully unpolarized  but massive case at nest-to-leading order (NLO)
was examined by \cite{Kaiser:2010zz}.

The scattering of muons off polarized electrons was  measured by  SMC collaboration at
CERN  in \cite{SpinMuonSMC:1997ixm}. At the request of SMC collaboration, the working tool 
for estimating theoretical uncertainty under experimental conditions was our code 
$\mu$ela  \cite{Bardin:1997nc} . 
The code for elastic polarized $\mu - e$ scattering was implemented at QED level 
with the possibility of taking into account the effects of longitudinal 
polarization 
in variables of the experiment with necessary cut-offs.

Recently, interest in the elastic 
muon-electron scattering has increased  due to high energy ($\mu$TRISTAN, KEK) and low energy (MUonE, CERN) experiments.

The new $\mu-e$ collider experiments $\mu$TRISTAN with the energies 
($E_{e^-}, E_{\mu^+}$) = (30 GeV, 1 TeV)
and the beam polarizations with $P_{e^-} = \pm 0.7$
and $P_{\mu^+} = \pm 0.8$ is currently under consideration at KEK \cite{Hamada:2022mua}.
 At such high energies polarised $\mu-e$ scattering has not been thoroughly investigated.

Realistic conditions of the MUonE experiment~\cite{Abbiendi:2016xup,Abbiendi:2677471} are: 
the energy of incoming muon $E_\mu=150$ GeV
in the laboratory system which corresponds to the center-of-mass system (c.m.s.) energy  
$s=m_e^2+m_\mu^2+2m_e E_\mu \approx$~(405~MeV)$^2$ (see also proposals ~\cite{Masiero:2020vxk},\cite{Dev:2020drf} for new physics searches).

Further development of the most advanced 
theoretical support at the NLO and partly NNLO levels by MC codes for this experiment is provided by three scientific groups:~(\cite{Alacevich:2018vez,CarloniCalame:2019mbo,CarloniCalame:2020yoz,Budassi:2021twh}), (\cite{Banerjee:2020rww,Ulrich:2020frs}), and (\cite{Fael:2019nsf,Balzani:2020yxg}).
A review of the current state of the theoretical support related to the proposed MUonE experiment and a discussion of further required steps can be found in~\cite{Banerjee:2020tdt}.

The main results of this work is an independent calculation of 
the complete one-loop electroweak radiation corrections (RCs) 
to elastic muon-electron scattering 
\bqa
\label{muplus}
\mu^+(p_1,\chi_1)\ +\ e^-(p_2,\chi_2)\ & \rightarrow & 
e^-(p_3,\chi_3)\ +\ \mu^+(p_4,\chi_4)\ \ (+ \gamma (p_5, \chi_5)),  \\
\label{muminus}
\mu^-(p_1,\chi_1)\ +\ e^-(p_2,\chi_2)\ & \rightarrow &
e^-(p_3,\chi_3)\ +\ \mu^-(p_4,\chi_4)\ \ (+ \gamma (p_5, \chi_5)),
\eqa
with 
arbitrary polarizations of initial particles 
($\chi$ corresponds the helicity of the particles).

We perform a comparison 
for  the  hard  bremsstrahlung  cross  section and
check the numerical dependence 
of Born and corrected cross sections on the muon and electron polarization degrees. 
We present results for low energy in  conditions
of the MUonE experiment.
To verify the effects of weak interactions we present  numerical results also 
for the c.m.s. energy $\sqrt{s}$ = 250~GeV. 
Here we also computed leading higher-order (h.o.) corrections of the 
${\cal O}(G_\mu^2)$ and ${\cal O}(G_\mu\alpha_s)$ orders through the parameters 
$\Delta \rho$ and $\Delta \alpha$. 
Our results
can be considered as a preliminary glance at electroweak and polarization 
effects 
a new high-energy muon-electron collider~\cite{Hamada:2022mua}.

This article is organized as follows.  Section~\ref{Sect2} contains preliminary remarks and the general notation. 
We describe the methodology for calculating polarized cross sections at the complete one-loop EW level 
within the helicity approach. Section~\ref{Sect3} gives a description of leading h.o. EW corrections 
in ${\cal O}(G_\mu^2)$ and ${\cal O}(G_\mu\alpha_s)$ orders.
In Section~\ref{Sect:results} 
we collect numerical results with various polarization degrees for total and differential cross sections as well as for relative corrections for high and low energies.
The last Section~\ref{Sect:Concl} contains discussion and conclusions.

\section{EW one-loop radiative corrections \label{Sect2}}

With the help of the computer system \SANC~\cite{Andonov:2004hi},
we have calculated the complete one-loop electroweak radiative
corrections to a wide class of processes, see the review~\cite{Bardin:2019zsp} and references therein.
Recently, the system has been upgraded~\cite{Bardin:2017mdd,Bondarenko:2020hhn}
in order to take into account possible longitudinal polarization of the initial particles.  
All calculations in the system can now be performed within the 
helicity amplitude formalism taking into account the initial and final state fermion masses. 
So the \SANC~ system provides a solid framework 
to access polarization effects at the one-loop level 
and study various effects, for example, the system allows one to separate the effects 
due to QED, weak radiative corrections, and some higher-order contributions;
to study radiative corrections in the $\alpha(0)$ and $G_\mu$ EW schemes.

A covariant amplitude (CA) corresponds to the result of the straightforward standard
calculation by means of the \SANC~ programs and procedures of {\it all} diagrams contributing 
to a given process at the tree (Born) and one-loop levels.
In the \SANC~ approach, we exploit the fact that the calculated one-loop form factors (FF) 
for the process $f \bar f  f \bar f \to 0$ can be turned to any other channel.
We describe the CA for this process decomposed on a massive basis with 
the so-called $\gamma$, $QQ$, $QL$, $LQ$, $LL$, $LD$, and $QD$ contributions.
It corresponds to six Dirac structures, see~\cite{Andonov:2002xc}.
They are labeled according to their structures.
We have the same one-loop FFs
for Bhabha scattering  $e^+ e^- \rightarrow e^- e^+$, 
$s$ channel $e^+ e^- \rightarrow \mu^- \mu^+(\tau^- \tau^+)$, 
elastic $\mu - e$ scattering  $\mu^\pm e^- \rightarrow e^- \mu^\pm$
and M{\o}ller scattering  $e^- e^- \rightarrow e^- e^-$.
Form factors  differ only by the corresponding permutations of their arguments $s,t,u$.
The description of their implementation for Bhabha scattering and $s$ channel annihilation processes
was presented in earlier articles~\cite{Bardin:2017mdd,Bondarenko:2020hhn}.

The complete one-loop cross section of the process can be split into four parts:
\bqa
\sigma^{\text{one-loop}} = \sigma^{\mathrm{Born}} + \sigma^{\mathrm{virt}}(\lambda)
+ \sigma^{\mathrm{soft}}(\lambda,\omega) + \sigma^{\mathrm{hard}}(\omega),
\eqa
where $\sigma^{\mathrm{Born}}$ is the Born cross section,
$\sigma^{\mathrm{virt}}$ is the contribution of virtual (loop) corrections,
$\sigma^{\mathrm{soft(hard)}}$ is the soft (hard) photon emission contribution
(the hard photon energy {$E_{\gamma} > \omega$}).
Auxiliary parameters $\lambda$ ("photon mass")
and $\omega$ are canceled out after summation.
Note that in calculations of one-loop RCs we can separate QED and pure 
weak interaction effects.

In \SANC, all one-loop contributions, i.e.,
the contributions of the virtual part as well as soft and hard bremsstrahlung radiation,
are realized within the helicity amplitude (HA) approach 
as in several other modern codes for theoretical support of experiments, see e.g.~\cite{Kilian:2007gr}.
In HA we keep the full dependence on the muon masses.
As an additional bonus of the HA approach, 
we implement a procedure to compute the cross section with any polarization of the initial
and final particles.

Typical {\tt SANC} results for FF are given in terms of 
only {\em scalar} Passarino-Veltman (PV) functions~\cite{Passarino:1978jh}.
In FF calculations at the one-loop level we keep all masses (without any approximation), 
but in this study we neglect contributions suppressed by the ratio $m_e^2/Q^2$
in the process cross section.

\subsection{Differential cross section}

To study the case of longitudinal polarization with degrees of $P_{e^+}$ and $P_{e^-}$, 
we calculate helicity amplitudes and make a formal application of equation ~(1.15) from~\cite{MoortgatPick:2005cw,Moortgat-Picka:2015yla}:
\begin{equation}
\sigma{(P_{e^+},P_{e^-})} = \frac{1}{4}\sum_{\chi_1,\chi_2}(1+\chi_1P_{e^+})(1+\chi_2P_{e^-})\sigma_{\chi_1\chi_2},
\label{eq1}
\end{equation}
where $\chi_{1(2)} = -1(+1)$ corresponds to the particle $i$ with left (right) helicity.

The virtual (Born) cross section of the $\mu^\pm e^- \to e^- \mu^\pm$ process can be written as follows:
\bqa
\frac{d\sigma^{\mathrm{virt(Born)}}_{ \chi_1 \chi_2}}{d\cos{\vartheta_e}}
= \pi\alpha^2\frac{1}{2s}|\mathcal{H}^{\mathrm{virt(Born)}}_{\chi_1 \chi_2}|^2,
\eqa
where
\bqa
|\mathcal{H}^{\rm virt(Born)}_{ \chi_1 \chi_2}|^2 =
\sum_{\chi_3,\chi_4} |\mathcal{H}^{\rm virt(Born)}_{\chi_1 \chi_2\chi_3\chi_4}|^2,
\eqa
$\vartheta_e$ is the angle between the initial (anti-)muon
$\mu^\pm$ and final electron $e^-$.

The soft photon bremsstrahlung contributions
are factorized at the Born cross section, they are
given below.

The cross section for the hard photon bremsstrahlung reads
\bqa
\frac{d\sigma^{\mathrm{hard}}_{{ \chi_1}{ \chi_2}{}{}{} }}{ds'
        d\cos{\theta_4 }d\phi_4
        d\cos{\theta_5}}
=\alpha^3\frac{s-s'}{128\pi s s'}
\frac{\sqrt{\lambda(s',m_\mu^2,m_e^2)}}{\sqrt{\lambda(s,m_\mu^2,m_e^2)}}
|\mathcal{H}^{\mathrm{hard}}_{{ \chi_1}{ \chi_2}{}{}{}} |^2,
\eqa
where $s'=(p_3+p_4)^2$, $\lambda(x,y,z)$ is the K\"all\'en (triangle) function,
and
\bqa
|\mathcal{H}^{\mathrm{hard}}_{ \chi_1 \chi_2{}{}{}} |^2 =
\sum_{\chi_3,\chi_4,\chi_5} |\mathcal{H}^{\mathrm{hard}}_{{ \chi_1}{ \chi_2}{ \chi_3}{ \chi_4}{\chi_5}} |^2.
\eqa
Here $\theta_5$ is the angle between 3-momenta of the photon and electron,
$\theta_4$ is the angle between 3-momenta of the (anti-)muon $\mu^{\pm}$ and photon in the rest
frame of the $(\mu^{\pm}e^-)$-compound, $\phi_4$ is the azimuthal angle of the 
(anti-)muon $\mu^{\pm}$ in the same frame.

\subsection{Helicity amplitudes for the Born and virtual part}

For both channels there are eight non-vanishing independent
HAs which depend on kinematic variables, coupling constants, and
six scalar FFs in the $LQD$-basis for $Z$ boson and one for $\gamma$ exchange. 

The HAs for $\mu^+$ channel have the following form:
\begin{eqnarray}
{\cal H}_{\mp\mp\mp\mp} &=&
    \frac{1}{t}  \Biggl\{
    s^- (2-\cpl r_s) \tffgg
\\ \nonumber &+&
  \chizt \Bigl[ s^- \left(
  -2 \cpl r_s \tffll+(2-\cpl r_s)(\tffqq+\tfflq+\tffql)-2\mmo^2\cmi(\tffld+\tffqd)
  \right)
  \\ \nonumber
  &\pm& \sqrt\lambda_{\mu e}    
  \left(-2 \cpl r_s\tffll+(2-\cpl r_s)\tfflq-(2+\cpl r_s)\tffql-2\cmi\mmo^2\tffld\right)
 \Bigr] \
 \Biggr\},
\\ \nonumber
{\cal H}_{\mp--\pm} &=&
    \sin\vartheta_e\frac{\mmo}{\sqrt{s}t}  \Biggl\{
    s^- \tffgg
       \\ \nonumber &+&
  \chizt \Bigl[ s^- \left(2 \tffll+\tffqq+\tfflq+\tffql-s^+ (\tffld+\tffqd)\right)
  \\ \nonumber
  &+& \sqrt\lambda_{\mu e}    
  \left(2 \tffll+\tfflq+\tffql-s^+\tffld\right)
  \Bigr] \
  \Biggr\},
\\ \nonumber          
{\cal H}_{\mp++\pm} &=&
   -\sin\vartheta_e\frac{\mmo}{\sqrt{s}t}  \Biggl\{
    s^- \tffgg
       \\ \nonumber &+&
  \chizt \Bigl[ s^- \left(2 \tffll+\tffqq+\tfflq+\tffql-s^+ (\tffld+\tffqd)\right)
  \\ \nonumber
  &-& \sqrt\lambda_{\mu e}    
  \left(2 \tffll+\tfflq+\tffql-s^+\tffld\right)
  \Bigr] \
  \Biggr\}.
  \\ \nonumber          
{\cal H}_{\mp\pm\pm\mp} &=&
    -\frac{\cmi}{t}  \Biggl\{
   s^-\tffgg
\\ \nonumber &+&
  \chizt \Bigl[ s^- 
 \left(2\tffll+\tffqq+\tfflq+\tffql-2\mmo^2(\tffld+\tffqd)\right)
  \\ \nonumber
  &\mp& \sqrt\lambda_{\mu e}    
  \left(2\tffll+\tfflq+\tffql-2\mmo^2\tffld\right)
 \Bigr] 
 \Biggr\},
\nonumber
\end{eqnarray}
where $\lambda_{\mu e}=\lambda(s,m^2_\mu,m^2_e)$.

The HAs for $\mu^-$ - channel have the following form:
\begin{eqnarray}
 {\cal H}_{\pm\pm\pm\pm} &=&-\frac{1}{t}
  \Biggl\{
  s^- (2-\cpl r_s) \tffgg 
  \\ \nonumber &+&
  \chizt  \Bigl[
   s^-  \left(4 \tffll +(2-\cpl r_s) (\tffqq+\tfflq+\tffql) -2\mmo^2\cmi (\tffld+\tffqd) \right)
  \\ \nonumber &\mp& \sqrt\lambda_{\mu e}\left(4 (\tffql+\tffll)+(2-\cpl r_s) (\tfflq-\tffql) -2 \mmo^2\cmi \tffld \right)
                  \Bigr]
  \Biggr\},
\\ \nonumber  
{\cal H}_{-\pm\pm+}
 &=&
    -\sin\vartheta_e\frac{\mmo}{t\sqrt{s}}\Biggl\{
        s^- \tffgg
        \\ \nonumber &+&
       \chizt \Bigl[
        s^-   \left(\tffqq+\tfflq+\tffql-s^+(\tffld+\tffqd)\right)
\\ \nonumber &\mp&       
\sqrt\lambda_{\mu e}      \left(\tfflq-\tffql-s^+\tffld \right)
                 \Bigr]  \Biggr\},
\\ \nonumber  
    {\cal H}_{+\pm\pm-}
 &=&
 \sin\vartheta_e\frac{\mmo}{t\sqrt{s}}\Biggl\{
       s^- \tffgg 
              \\ \nonumber &+&
       \chizt \Bigl(
        s^- \left[\tffqq+\tfflq+\tffql-s^+(\tffld+\tffqd)\right]
         \\ \nonumber &\mp&    
       \sqrt\lambda_{\mu e}(\tfflq-\tffql-s^+\tffld)
                         \Bigr)\Biggr\},
\\ \nonumber  
{\cal H}_{\pm\mp\mp\pm} &=&-\frac{\cmi}{t}\Biggl\{
        s^- \tffgg 
               \\ \nonumber &+&
      \chizt
      \Bigl[     
s^-( \tffqq+\tfflq+\tffql-2\mmo^2(\tffld+\tffqd))
      \\&\pm&  
       \sqrt\lambda_{\mu e}(\tfflq-\tffql-2\mmo^2\tffld)\Bigr]\Biggr\}
       \nonumber
\end{eqnarray}
with
$s^- = s - \mmo^2, \quad s^+ = s + \mmo^2,$ 
and $\chi_{\sss{Z}}(t)$ being the $Z/\gamma$ propagator ratio:
\bqa
\chizt &=& \frac{1}{4\sw\cw}
\frac{\ds t}{\ds{t - \mz^2}}\,.
\label{propagators}
\eqa

For both channels the other six HAs are expressed through the above ones as follows:
${\cal H}_{---+} = {\cal H}_{+---}$, 
${\cal H}_{--++} = {\cal H}_{++--}$,
${\cal H}_{-+--} = {\cal H}_{--+-}$,
${\cal H}_{-+-+} = {\cal H}_{+-+-}$,
${\cal H}_{-+++} = {\cal H}_{+++-}$,
${\cal H}_{+-++} = {\cal H}_{++-+}$.

The helicity indices denote the signs of the fermion spin projections to their momenta $p_1, p_2, p_3, p_4$, 
respectively. The notation
\begin{equation}
c^{\pm}=1 \pm \cos{\vartheta_{e}}
\end{equation}
was introduced. The electron scattering angle $\vartheta_{e}$
is related to the Mandelstam variables $t = (p_--p_4)^2$ and $u = (p_2-p_3)^2$:
\bqa
\cos\vartheta_\mu &=& \bigg[s(u-t) - (m^2_\mu-m^2_e)^2\bigg]/\lambda_{\mu e}.
\eqa

Note that the {\it tilded} form factors absorb couplings, which leads to compactification 
of the formulae for the HAs.
The expressions for the {\it tilded} form factors are
\begin{eqnarray}
\tffgg \;\; &=& \qel\qmo \ffgg,
\label{tffs}\\
\tffll &=& \ael\amo \ffll,
\nonumber\\ 
\tfflq &=& \ael\vmamo  \fflq,
\nonumber\\
\tffql &=& \vmael\amo  \ffql,
\nonumber\\
\tffqq &=& \vmael\vmamo \ffqq,
\nonumber\\
\tffld &=& \ael\amo    \ffld,
\nonumber\\
\tffqd &=&\vmael\amo   \ffqd,
\nonumber\\
\end{eqnarray}
with the coupling constants
\bqa
I^{(3)}_f\,,\quad \sigma_f = v_f + a_f\,,\quad \delta_f = v_f - a_f\,,
\quad \stw=\frac{e}{g}\,,\quad 
\ctw=\frac{\mw}{\mz}.
\eqa

In order to get HAs for the Born level, one should set
$F_{\sss \gamma,LL,LQ,QL,QQ}=1$ and $F_{\sss LD,QD}=0$.

\subsection{The soft photon emission bremsstrahlung}
\label{Sect22}

The soft photon contribution contains the infrared divergences and 
compensates the corresponding divergences of the one-loop virtual QED corrections.
This soft photon brems\-strahlung correction can be calculated analytically, and 
it is factorized at the Born cross section. 
In this case, the polarization dependence is contained only in the Born cross 
section $\sigma^{\rm Born}$.

\bqa
\sigma^{{\rm soft},{\rm \mu-leg}} &=&
       -\qmo^2  \frac{\alpha}{\pi} \sigma^{\rm Born}\Biggl\{ 
       \lnlaom+\frac{s+\rmoms}{s-\rmoms}\ln\left(\rms\right)    
   -\frac{\ds 1+2 \rmt}{\ds\sqrt{1+4\rmt}}
      \Biggl[ \ln(a_{14})\lnlaom    
   \\ &&  
        +\Litwo\left(1-\frac{a_{14} \xmo}{v_{14}}\right)    
      -\Litwo\left(1-\frac{\xmo}{v_{14}}\right)
      +\Litwo\left(1-\frac{a_{14}}{v_{14}\xmo}\right)
      -\Litwo\left(1-\frac{1}{v_{14}\xmo}\right) \Biggr] \Biggl\},
      \nonumber \\
\sigma^{{\rm soft},{\rm ifi}} &=&
      -\qel \qmo \frac{\alpha}{\pi} \sigma^{\rm Born}\Biggl\{
        2 \ln(\xmo \xel) \lnlaom
        + \Litwo\left(1-\xmo^2\right)+\Litwo\left(1-\xel^2\right)
\nll &&        
        -\Litwo\left(\ds 1-\frac{1}{\xmo^2}\right)-\Litwo\left(1-\frac{1}{\xel^2}\right) 
        - \ln\left(a_{13}\right) \lnlaom
\nll &&        
        -\Litwo\left(\ds 1-\frac{a_{13} \xmo}{v_{13}}\right)
      -\Litwo\left(\ds 1-\frac{\xel}{v_{13}}\right)
        -\Litwo\left(\ds 1-\frac{a_{13}}{v_{13}\xmo}\right)+\Litwo\left(1-\frac{1}{v_{13}\xel}\right) 
\nll &&        
        - \ln\left(a_{24}\right) \lnlaom
        -\Litwo\left(\ds 1-\frac{a_{24}\xel}{v_{24}}\right)
\nll &&        
        +\Litwo\left(\ds 1-\frac{\xmo}{v_{24}}\right)
        -\Litwo\left(\ds 1-\frac{a_{24}}{v_{24}\xel}\right)
        +\Litwo\left(\ds 1-\frac{1}{v_{24}\xmo}\right)      
       \Biggr\},
\nonumber \\
\sigma^{{\rm soft},{\rm e-leg}} 
&=& -\qel^2 \frac{\alpha}{\pi} \sigma^{\rm Born}\Biggl\{
   \lnlaom+\ln\left({\ds \frac{m_e^2}{s(1-\rms^2)}}\right)
   - \ln\left(a_{23}\right) \lnlaom
\nll &&   
   -\Litwo\left(\ds 1-\frac{a_{23}\xel}{v_{23}}\right)
   +\Litwo\left(\ds 1-\frac{\xel}{v_{23}}\right)
   -\Litwo\left(\ds 1-\frac{a_{23}}{v_{23}\xel}\right)
   +\Litwo\left(\ds 1-\frac{1}{v_{23}\xel}\right) 
                                   \Biggr\}.
\nonumber \\
\xmo &=& \frac{\sqs}{\mmo}, \quad \xel = \frac{\sqs}{\mel}\left(1-\rms\right),
\nll
 a_{14} &=& \frac{1}{2\rmoms}\left(2 \rmoms+\ts+\sqrt{\ts^2+4 \rmoms \ts}\right),\quad
 v_{14} =\frac{\mmo}{\sqs (1+\rms)} \left(a_{14}+1\right),
 \nll
      a_{23} &=& \frac{\ts}{\relms},\quad
      v_{23} = \frac{\mel}{\sqs(1-\rms)} (a_{23}+1),
\nll
      a_{13} &=& \frac{(\us+\rmoms)}{\mel\mmo}, \quad
      v_{13} = \frac{a_{13}^2}{\sqs\left(a_{13}(1+\rms)/\mmo-(1-\rms)/\mel\right)},
\nll
      a_{24} &=& a_{13},\quad v_{24} = \frac{a_{24}^2}{\sqs\left( a_{24}(1-\rms)/\mel-(1+\rms)/\mmo\right)},
\nll      
r_{I}&=&\frac{\rmoms}{I},\quad I=s,t.
\eqa

\subsection{Helicity amplitudes for hard photon bremsstrahlung}

We use the \SANC~ spinor amplitude module for the $e^+ e^- l^+ l^- \gamma \to 0$
($\sum p_i = 0$) process in any of the $s$, $t$ or $u$ channels, 
where $0$ stands for {\em vacuum} and all masses are not neglected.

The full expression for the photon bremsstrahlung amplitude of the process under investigation 
can be divided into two terms
\bqa
\mathrm{A}_{\ldots \chi_i \ldots} =  2\sqrt{2} \left(Q_{\mu} A^{\mu}_{\ldots \chi_i \ldots} 
+ Q_e A^e_{\ldots \chi_i \ldots}\right).
\eqa
Each term corresponds to a gauge-invariant diagram subset: $A^{\mu}$ is the amplitude for radiation
off the muon line (MSR) and $A^e$ for radiation off the electron line (ESR).

There exists the following crossing symmetry relation between MSR and ESR amplitudes:
\bqa
&&    A^{\mu} \hel{ \chi_1}{ \chi_4}{ \chi_3}{ \chi_2}{\chi_5}(p_1, p_4, p_3, p_2,p_5)
    \\
    &&  =  A^e \hel{ \chi_3}{ \chi_2}{ \chi_1}{ \chi_4}{\chi_5}(p_3, p_2, p_1, p_4,p_5)
~~\mbox{and}~~ m_\mu \leftrightarrow m_e.
\nonumber    
\eqa

The explicitly gauge invariant form of the amplitude is obtained in~\cite{Bondarenko:2020hhn} and implemented 
as a \SANC~ module:
\begin{gather}
\begin{aligned}
\sqrt{2} A^{e}_{\chi_1\chi_2\chi_3\chi_4\chi_5} &= 
\dfrac{\Tr[\slashed{p}_1 \slashed{p}_2\mathbf{F}_5] }{z_{1} z_{2}}
	\bar{v}_1 \slashed{e}_{34} u_2 \\
	&-\dfrac{\bar{v}_1\mathbf{F}_5 \slashed{e}_{34} u_2}{z_{1}}
	-\dfrac{\bar{v}_1 \slashed{e}_{34} \mathbf{F}_5 u_2}{z_{2}},
\end{aligned}
\end{gather}
with abbreviations $z_{i} =2 p_i\cdot p_5$, $u_i \equiv u^{\chi_i}(p_i)$  etc.

The polarization vector of a real photon appears only in the combination
$\mathbf{F}_5 = p_5^{\mu}\varepsilon_5^{\nu}\sigma_{\mu\nu}$.
This is the familiar Maxwell bi-vector which is gauge invariant. 
We introduce also abbreviations for the following combinations of propagators 
and couplings constants:
\begin{gather}
\begin{aligned}
  \slashed{e}_{34} &= \dfrac{1}{2}\sum_{a,b=\pm1}\mathcal{D}^{ab}  (\bar{v}_3 \gamma^{\mu}\gamma_{b} u_4)\gamma_{\mu}\gamma_{a},  
\\
\mathcal{D}^{ab} &=\dfrac{Q_e Q_l}{s'} + \dfrac{g_e^{a}g_{l}^{b}}{s'-M_Z^2 + iM_Z\Gamma_Z}, 
 	\end{aligned}
\end{gather}
where $g_l^{\pm}$ are the chiral couplings of the leptons $l$ to the vector boson $Z$.

We work in the chiral representation of gamma-matrices and exploit Weyl spinors.
Our notation is consistent with~\cite{Badger:2005zh,Badger:2005jv,MAITRE2008501}. 
Below we use the following notation for the decomposition of Dirac spinors into Weyl ones:
\begin{gather}
\begin{aligned}
\slashed{p} &= 
\begin{pmatrix}
   & p_{A\dot{B}} \\
p^{\dot{A}B} &  
\end{pmatrix} = 
\begin{pmatrix}
   & \check{p} \\
\hat{p} &  
\end{pmatrix} 
,&
u  &= \begin{pmatrix}
 u_{A} \\   u^{\dot{A}} 
\end{pmatrix}= \begin{pmatrix}
 \SpIA{u} \\   \SpIB{u} 
\end{pmatrix},
\end{aligned}
\nonumber\\
\begin{aligned}
\bar{u}  &= \begin{pmatrix}
 \bar{u}^{A} ,&   \bar{u}_{\dot{A}} 
\end{pmatrix}= \begin{pmatrix}
 \SpAI{\bar{u}} ,&   \SpBI{\bar{u}} 
\end{pmatrix},
\\
\mathbf{F} &=\begin{pmatrix}
  \mathbf{F}_A{}^{B} & \\ &  \mathbf{F}^{\dot{A}}{}_{\dot{B}}
\end{pmatrix}=\begin{pmatrix}
\check{\mathbf{ F}} & \\ & \hat{\mathbf{F}}
\end{pmatrix}.
\end{aligned}
\end{gather}
Application of the Fiertz identities to the Pauli matrices yields
\begin{gather}
\begin{aligned}
 \hat{e}_{34} &=  \SpIB{\bar{v}_3} \mathcal{D}^{++} \SpAI{u_4} + \SpIB{u_4} \mathcal{D}^{+-}\SpAI{\bar{v}_3}
,\\
 \check{e}_{34} &= \SpIA{u_4} \mathcal{D}^{-+} \SpBI{\bar{v}_3} + \SpIA{\bar{v}_3} \mathcal{D}^{--}\SpBI{u_4} 
 .
\end{aligned}
\end{gather}

The HAs are not Lorentz-invariant objects (they are changed by boosts transverse to a momentum ray) and
thus rudimentally depend on an experimental setup. However, one expects that the entire physical content of a
reaction should depend only on a relative configuration of particles by analogy with the rigid
body dynamics in a rotating reference frame. This type of description usually appears to be most
economic one. 

In order to factor-out all information related to an experiment configuration, we must build a spin
basis in terms of the given problem's momenta. Our investigations show that one of the most 
economic choices is to put the polarization vector of fermion $n_i$ with $i=1,\ldots,4$ 
into the same 2-plane with its momentum $p_i$ and momentum of photon $p_5$. 
Each 2-plane contains two light-like vectors:
photon momentum and the other one denoted by $k_i$.
The explicit expressions for $k_i$ can be found in~\cite{Bondarenko:2018sgg}.
Then in the photon basis we have
\begin{align}
 \SpIA{u_{i}^{+} } &= \SpIA{v_{i}^{-}} = \SpIA{\bar{u}_{i}^{-} } = \SpIA{\bar{v}_{i}^{+}}= \SpIA{k_i} \equiv \SpIA{i} 
,\\
 \SpIB{u_{i}^{-} } &= \SpIB{v_{i}^{+}} = \SpIB{\bar{u}_{i}^{+}} = \SpIB{\bar{v}_{i}^{-}} = \SpIB{k_i} \equiv \SpIB{i}
,\nonumber\\
 \SpIB{u_{i}^{+} } &= -\SpIB{v_{i}^{-}} = -\SpIB{\bar{u}_{i}^{-}} = \SpIB{\bar{v}_{i}^{+}} = -\SpIB{5}\varsigma_i^{*}
,\nonumber\\
 \SpIA{u_{i}^{-} } &= -\SpIA{v_{i}^{+}} = -\SpIA{\bar{u}_{i}^{+}} = \SpIA{\bar{v}_{i}^{-}} =  \SpIA{5}\varsigma_i,
\nonumber\end{align}
with $\varsigma_i= m_i/\SpAIA{i}{5}$, where we identify $k_5\equiv p_5$.

A Maxwell bi-vector has a factorized form in the spinor notation:
$ \check{\mathbf{F}}_5^{+} = \sqrt{2}\;\SpIAoAI{5}{5}$,  
$\hat{\mathbf{F}}_5^{-}  = \sqrt{2}\;\SpIBoBI{5}{5}$, 
$\check{\mathbf{F}}_5^{-} = \hat{\mathbf{F}}_5^{+}=0$,
which allows us to organize the amplitude in terms of blocks
\begin{gather}
\begin{aligned}
 A^{e}_{\chi_1\chi_2\chi_3\chi_4\chi_5} &= 
\mathcal{S}_{\chi_5} \mathcal{B}_{\chi_1\chi_2\chi_3\chi_4}
\\
	&-\mathcal{C}^1_{\chi_1\chi_5} \mathcal{G}^2_{\chi_2\chi_3\chi_4\chi_5}
	-\mathcal{C}^2_{\chi_2\chi_5} \mathcal{G}^1_{\chi_1\chi_3\chi_4\chi_5} 
,
\end{aligned}
\end{gather}
where
\begin{gather}
\begin{aligned}
\mathcal{B}_{\chi_1\chi_2\chi_3\chi_4}  
= \SpBIIA{\bar{v}_1}{\hat{e}_{34}}{u_2} + \SpAIIB{\bar{v}_1}{\check{e}_{34}}{u_2}
,
\end{aligned}
\end{gather}
\begin{gather*}
\begin{aligned}
\mathcal{G}^1_{\chi_1\chi_3\chi_4\pm} &
=\begin{bmatrix}
\SpBIIA{\bar{v}_1}{ \hat{e}_{34}}{5}
\nonumber\\
 \SpAIIB{\bar{v}_1}{ \check{e}_{34}}{5}
\end{bmatrix} 
,&
\mathcal{G}^2_{\chi_2\chi_3\chi_4\pm} & 
=\begin{bmatrix}
\SpAIIB{5}{\check{e}_{34}}{u_2}
\nonumber\\
 \SpBIIA{5}{\hat{e}_{34}}{u_2}
\end{bmatrix},
\end{aligned}
\nonumber\\
\begin{aligned}
\mathcal{S}_{\chi_5} &
=-\begin{bmatrix}
  \dfrac{ \SpBIB{1}{2}  }{ \SpBIB{1}{5}  \SpBIB{2}{5} }
,&
 \dfrac{ \SpAIA{1}{2}  }{\SpAIA{1}{5}  \SpAIA{2}{5} }
\end{bmatrix}, 
\end{aligned}
\nonumber\\
\begin{aligned}
\mathcal{C}^1_{\chi_1\chi_5} &	
=\begin{bmatrix}
 \SpBIB{5}{1}  &   \nonumber\\
   &  \SpAIA{5}{1} 
\end{bmatrix}^{-1}
,&
\mathcal{C}^2_{\chi_2\chi_5} &
= \begin{bmatrix}
 \SpBIB{2}{5}  &  \nonumber\\
  &  \SpAIA{2}{5} 
\end{bmatrix}^{-1}
.
\nonumber
\end{aligned}
\end{gather*}

We are going to evaluate the amplitude only for positive photon helicity
because the case of negative one can be easily obtained with the help of the CP symmetry: 
\bqa
    {A}\hel{\chi_1}{\chi_2}{\chi_3}{\chi_4}{-} =
	-{\chi_1}{\chi_2}{\chi_3}{\chi_4}
	{A}^{*} \hel{-\chi_1}{-\chi_2}{-\chi_3}{-\chi_4}{+},
        \nonumber\\
\eqa
with ``+'' $\leftrightarrow$ ``--'' in $\mathcal{D}^{ab}$.
Below we give all amplitudes with positive photon helicity:
\begin{gather*}
\begin{aligned}
A^e_{{-}{-}{-}{+}{+}} &= -\mathcal{S}_{+}\SpAIA{4}{5}\big( \mathcal{D}^{++}\SpBIB{3}{1} \varsigma_{2} 
+ \mathcal{D}^{-+}\SpBIB{3}{2}\varsigma_{1} \big)  ,
\nonumber\\
A^e_{{-}{-}{+}{-}{+}} &= -\mathcal{S}_{+}\SpAIA{3}{5}\big( \mathcal{D}^{+-}\SpBIB{4}{1} \varsigma_{2} 
+ \mathcal{D}^{--}\SpBIB{4}{2}\varsigma_{1} \big)  ,
\nonumber\\
A^e\labhel{}{-}{+}{-}{-}{+} &= \phantom{-}\mathcal{S}_{+}\SpAIA{2}{5}\Big(
\mathcal{D}^{++}\SpBIB{3}{1}\varsigma_{4} + \mathcal{D}^{+-}\SpBIB{4}{1}\varsigma_{3}
\Big),
\nonumber\\
A^e\labhel{}{+}{-}{-}{-}{+} &=\phantom{-}\mathcal{S}_{+}\SpAIA{1}{5}\Big(
\mathcal{D}^{-+}\SpBIB{3}{2}\varsigma_{4} + \mathcal{D}^{--}\SpBIB{4}{2}\varsigma_{3}
\Big),
\nonumber\\
A^e\labhel{}{-}{-}{+}{+}{+} &= \mathcal{S}_{+}  
\SpBIB{5}{1}\Big( 
\mathcal{D}^{+-}\SpAIA{3}{5}\varsigma_{2}\varsigma^{*}_{4}
+\mathcal{D}^{++}\SpAIA{4}{5}\varsigma_{2}\varsigma^{*}_{3}
 \Big)
\nonumber\\
 &+\mathcal{S}_{+}\SpBIB{5}{2}\Big(
  \mathcal{D}^{--}\SpAIA{3}{5}\varsigma_{1}\varsigma^{*}_{4}
+ \mathcal{D}^{-+}\SpAIA{4}{5}\varsigma_{1}\varsigma^{*}_{3}
\Big) ,
\nonumber\\
A^e\labhel{}{+}{+}{-}{-}{+} &= \mathcal{S}_{+} 
\SpAIA{1}{5}\Big(
  \mathcal{D}^{-+}\SpBIB{5}{3}\varsigma^{*}_{2}\varsigma_{4}
+ \mathcal{D}^{--}\SpBIB{5}{4}\varsigma^{*}_{2}\varsigma_{3}
\Big)
\nonumber\\
 &+ \mathcal{S}_{+} \SpAIA{2}{5}\Big(\mathcal{D}^{++}\SpBIB{5}{3}\varsigma^{*}_{1}\varsigma_{4}
                   + \mathcal{D}^{+-}\SpBIB{5}{4}\varsigma^{*}_{1}\varsigma_{3}
\Big) ,
\end{aligned}
\end{gather*}
\begin{gather*}
\begin{aligned}
 A^e\labhel{}{-}{+}{-}{+}{+} &=\mathcal{S}_{+}\Big(
\mathcal{D}^{++}\SpBIB{3}{1}\SpAIA{2}{4} 
-   \mathcal{D}^{-+}\SpAIA{4}{5}\SpBIB{5}{3}\varsigma_{1}\varsigma^{*}_{2}
\nonumber\\
&-  \mathcal{D}^{+-}\SpAIA{2}{5}\SpBIB{5}{1}\varsigma_{3}\varsigma^{*}_{4}
\Big) 
+ \mathcal{C}^{2}_{++}\mathcal{D}^{++}\SpBIB{3}{1}\SpAIA{4}{5},
\nonumber\\ 
A^e\labhel{}{-}{+}{+}{-}{+} &=\mathcal{S}_{+}\Big(
\mathcal{D}^{+-}\SpBIB{4}{1}\SpAIA{2}{3} 
-   \mathcal{D}^{--}\SpAIA{3}{5}\SpBIB{5}{4}\varsigma_{1}\varsigma^{*}_{2}
\nonumber\\
&-  \mathcal{D}^{++}\SpAIA{2}{5}\SpBIB{5}{1}\varsigma^{*}_{3}\varsigma_{4}
\Big) 
+ \mathcal{C}^{2}_{++}\mathcal{D}^{+-}\SpBIB{4}{1}\SpAIA{4}{5},
\nonumber\\
 A^e\labhel{}{+}{-}{-}{+}{+} &=\mathcal{S}_{+}\Big(
\mathcal{D}^{-+}\SpBIB{3}{2}\SpAIA{1}{4} 
-   \mathcal{D}^{++}\SpAIA{4}{5}\SpBIB{5}{3}\varsigma^{*}_{1}\varsigma_{2}
\nonumber\\
&-  \mathcal{D}^{--}\SpAIA{1}{5}\SpBIB{5}{2}\varsigma_{3}\varsigma^{*}_{4}
\Big) 
+ \mathcal{C}^{1}_{++}\mathcal{D}^{-+}\SpBIB{3}{2}\SpAIA{4}{5}, 
\nonumber\\
 A^e\labhel{}{+}{-}{+}{-}{+} &=\mathcal{S}_{+}\Big(
\mathcal{D}^{--}\SpBIB{4}{2}\SpAIA{1}{3} 
-   \mathcal{D}^{+-}\SpAIA{3}{5}\SpBIB{5}{4}\varsigma^{*}_{1}\varsigma_{2}
\nonumber\\
&-  \mathcal{D}^{-+}\SpAIA{1}{5}\SpBIB{5}{2}\varsigma^{*}_{3}\varsigma_{4}
\Big) 
+ \mathcal{C}^{1}_{++}\mathcal{D}^{--}\SpBIB{4}{2}\SpAIA{3}{5}, 
\nonumber\\
 A^e\labhel{}{-}{+}{+}{+}{+} &= - \mathcal{S}_{+}\SpBIB{5}{1}\Big( 
   \mathcal{D}^{+-}\SpAIA{2}{3} \varsigma^{*}_{4} +  \mathcal{D}^{++}\SpAIA{2}{4} \varsigma^{*}_{3}
\Big) 
\nonumber\\
&- \mathcal{C}^{2}_{++}\SpBIB{5}{1}\Big( 
   \mathcal{D}^{+-}\SpAIA{3}{5} \varsigma^{*}_{4} +  \mathcal{D}^{++}\SpAIA{4}{5} \varsigma^{*}_{3}
\Big) ,
\nonumber\\
 A^e\labhel{}{+}{-}{+}{+}{+} &= - \mathcal{S}_{+}\SpBIB{5}{2}\Big( 
   \mathcal{D}^{--}\SpAIA{1}{3} \varsigma^{*}_{4} +  \mathcal{D}^{-+}\SpAIA{1}{4} \varsigma^{*}_{3}
\Big) 
\nonumber\\
&- \mathcal{C}^{2}_{++}\SpBIB{5}{2}\Big( 
   \mathcal{D}^{--}\SpAIA{3}{5} \varsigma^{*}_{4} +  \mathcal{D}^{-+}\SpAIA{4}{5} \varsigma^{*}_{3}
\Big),
\nonumber\\
 A^e\labhel{}{+}{+}{-}{+}{+} &= \mathcal{S}_{+}\SpBIB{5}{3}\Big( 
   \mathcal{D}^{-+}\SpAIA{1}{4} \varsigma^{*}_{2} +  \mathcal{D}^{++}\SpAIA{2}{4} \varsigma^{*}_{1}
\Big) 
\nonumber\\
&+ \SpAIA{4}{5}\SpBIB{5}{3}\Big(\mathcal{C}^{1}_{++} \mathcal{D}^{-+} \varsigma^{*}_{2} 
+  \mathcal{C}^{2}_{++} \mathcal{D}^{++} \varsigma^{*}_{1}
\Big) , 
\nonumber\\
 A^e\labhel{}{+}{+}{+}{-}{+} &= \mathcal{S}_{+}\SpBIB{5}{4}\Big( 
   \mathcal{D}^{--}\SpAIA{1}{3} \varsigma^{*}_{2} +  \mathcal{D}^{+-}\SpAIA{2}{3} \varsigma^{*}_{1}
\Big) 
\nonumber\\
&+ \SpAIA{3}{5}\SpBIB{5}{4}\Big(\mathcal{C}^{1}_{++} \mathcal{D}^{--} \varsigma^{*}_{2} 
+  \mathcal{C}^{2}_{++} \mathcal{D}^{+-} \varsigma^{*}_{1}
\Big) ,
\\
 A^e\labhel{}{+}{+}{+}{+}{+} &=  A^e\labhel{}{-}{-}{-}{-}{+} =0.
\end{aligned}
\end{gather*}

To obtain HA $\mathcal{H}$ with a definite set of helicities, the basis-transformation
matrices $ C_{\xi_i}^{\phantom{a}\chi_i}$ should be applied independently for each index $\chi$
of external particles whose polarization is not averaged:
\bqa
\mathcal{H}_{... \xi_i ...} \hspace*{-1mm}= 
C_{\xi_1}^{\phantom{a}\chi_1} \hspace*{-2mm}\dots C_{\xi_4}^{\phantom{a}\chi_4}
           \mathrm{A}_{... \chi_i ...}
\eqa
Explicit expressions for the matrices $C$ can be found in~\cite{Schwinn:2007ee}
and for our special case in~\cite{Bondarenko:2018sgg}.
Geometrically, they realize a Wigner rotation of the spin axis~\cite{Wigner:1957ep,Wigner:1964zf}.

\section{\label{Sect3} Leading higher-order electroweak effects}

\subsection{The $\rho$ parameter}

The electroweak parameter $\rho$, introduced by Veltman \cite{Veltman:1977kh}, 
measures the relative strength of charged and neutral currents.
The $\rho$ parameter is defined as the ratio
of a neutral current amplitude to
a charged current one at the zero momentum transfer,
see for example~\cite{Fleischer:1993ub}:
\begin{equation}
\rho=\frac{G_{NC}(0)}{G_{CC}(0)}=\frac{1}{1-\Delta\rho}\,,
\label{def-rho}
\end{equation} 
where $G_{CC}(0)=G_\mu$ is the Fermi constant defined from the $\mu$-decay width,
and the quantity $\Delta\rho$ is treated perturbatively
\begin{equation}
\Delta\rho=\Delta\rhoone+\Delta\rhotwo+\ldots
\end{equation}

Expanding (\ref{def-rho}) up to quadratic terms $\Delta\rho^2$, we have
\begin{equation}
\rho=1+\Delta\rho+\Delta\rho^2\,.
\label{rho-upto-quadro}
\end{equation}
The leading in $G_\mu m_t^2$ NLO EW contribution to $\Delta\rho$
is explicitly given by
\begin{equation}
\Delta\rhoone\Big|^{G_\mu}=3\xt=\frac{3\sqrt{2}G_\mu m_t^2}{16\pi^2}\,.
\label{def-rho-one}
\end{equation}

A large group of dominant radiative corrections can
be absorbed into the shift of the $\rho$ parameter from its
lowest order value $\rho_{\mathrm{Born}} = 1$.
The major contributions of these groups are:
\begin{eqnarray}
  \Delta\rho = 
    \Delta\rho_{X_t}                          
  + \Delta\rho_{\alpha\alpha_s}     
  + \Delta\rho_{X_t\alpha_s^2} 
 +\ldots
 \end{eqnarray}
We follow the 	
prescription introduced in  Refs.~\cite{Fleischer:1993ub,Fleischer:1994cb}
and later well described in Refs~\cite{Bardin:280836,Bardin:1999yd}.

At the two-loop level, the quantity $\Delta\rho$ contains two contributions:
\begin{equation}
\Delta\rho=N_c\xt\,\left[1+\rhotwo\,\left(\mh^2/\mt^2\right)\,\xt\,\right]\,
            \left[1-\frac{2\alpha_s(\mz^2)}{9\pi}(\pi^2+3)\right].
\end{equation}
They consist of the following:

i) the two-loop EW part at ${\cal O}(G_\mu^2)$, the second term in the first square 
brackets~\cite{Barbieri:1992nz,Fleischer:1993ub,Fleischer:1994cb}
with $\rhotwo$ given in Eq.~(12) of~\cite{Fleischer:1994cb}
(actually, after the Higgs boson discovery
and determination of its mass, it has become
sufficient to use the low Higgs mass asymptotics, Eq.~(15), of~\cite{Fleischer:1994cb}); 

ii) the mixed two-loop EW$\otimes$QCD at ${\cal O}(G_\mu\alpha_s)$,
the second term in the second square 
brackets, see Ref.~\cite{Djouadi:1987gn,Djouadi:1987di} for further details.

\subsection{Implementation of the $\Delta\alpha$ and $\Delta\rho$ parameters}

The leading higher-order effects are usually parameterized by 
$\Delta\alpha$ and $\Delta\rho$ and their two-loop contributions can be included
in a straightforward way.

The corrections induced by the running of $\alpha$ can be included 
by resummation the NLO EW $\Delta\alpha$ parameter
\bqa
\alpha(Q^2) = \frac{\alpha(0)}{1-\Delta\alpha(Q^2)}.
\label{alphat}
\eqa
Here $Q^2$ is the scale which characterizes the evolution of the EW coupling.
In the case of elastic $\mu - e$ scattering, $Q^2 = t$ where $t$ is the Mandelstam variable.

To implement the $\Delta\rho$ parameter, we introduce
intermediate vector boson propagators 
$\sim 1/(Q^2+M^2_V)$,
into Eq.~(\ref{def-rho}) and derive the following definition:
\begin{equation}
\rho=\frac{\mw^2}{\cwb\mz^2}\,,
\label{def-cwb}
\end{equation}
where we have introduced a new parameter $\cwb$ to distinguish from the usual $\cw$ 
for which we maintain the meaning
$\cw=\mw^2/\mz^2$ to be valid to all perturbative orders. At the lowest order (LO)
\begin{equation}
\rho^{(0)}=\frac{\mw^2}{\cw\mz^2}=1\,.
\end{equation}
From Eq.(\ref{def-cwb}) we have:
\begin{equation}
\cwb=\frac{\mw^2}{\rho\mz^2} = (1-\Delta\rho)\,\cw,\quad\quad
\swb=\sw+\Delta\rho\,\cw.
\label{cwb}
\end{equation}

Equations~(\ref{alphat}) and~(\ref{cwb}) allow us to introduce 
the leading higher-order effects via the following replacements:
\bqa
&&
\alpha(0) \to \alpha(t) = \alpha(0) 
\bigl[ 1+\Delta\alpha(t) + \Delta\alpha(t)^2 \bigr],
\\
&&\sw\to\swb\equiv\sw (1+\frac{\cw}{\sw}\Delta\rho)
\quad
\cw\to\cwb\equiv 1-\swb=(1-\Delta\rho)\,\cw\,
\label{eq:replacements}
\eqa
into the Born cross section with $\alpha(0)$ EW coupling.
The obtained recipe allows one to reproduce correctly the terms up to
${\cal O}(\Delta\rho^2)$ and ${\cal O}(\Delta\alpha(t)\Delta\rho)$
(\cite{Consoli:1989fg},\cite{Consoli:1989pc}, and \cite{Fleischer:1993ub}).

In the SANC we use replacements~(\ref{eq:replacements}) 
at the level of form factors. There are five Born-like form factors:
one for $\gamma$ exchange ${\tffgg}$ that is proportional only to the EW coupling 
and four for $Z$ exchange: $\tffll, \tfflq, \tffql, \tffqq$ that are all proportional to 
the EW coupling and to the factor $(\sw\cw)^{-1}$ from $\chizt$~Eq.~(\ref{propagators}).
Also each $Q$ form factor has an additional $\sw$,~Eq.~(\ref{tffs}). 

We consider the replacement for the form factors in the $\alpha(0)$ and $G_\mu$ EW schemes separately.

$\bullet$ {the $\alpha(0)$ EW scheme}

Let us consider the replacement for the $\tffll$ form factor due to all factors
listed above:
\bqa
&&\alpha(0) \frac{1}{\sw\cw} \tffll \to
\alpha(t) \frac{1}{\swb\cwb} \tffll = 
\\
&&=\alpha(0) \frac{1}{\sw\cw} \bigl[ 1 + \Delta\alpha(t) + \Delta\alpha(t)^2 \bigr]
\bigl[ 1 - \frac{\cw}{\sw}\Delta\rho + \frac{\ctw^4}{\stw^4} \Delta\rho^2 \bigr]
\bigl[ 1 + \Delta\rho + \Delta\rho^2 \bigr].
\eqa
Removing the common constant from the left and right sides, the leading two-loop form factors read
\bqa
\tffgg  &\to& 1+\Delta\alpha(t)+\Delta\alpha(t)^2,         \nll
\tffll  &\to& (1+\Delta\alpha(t)+\Delta\alpha(t)^2) (1 + \Delta\rho+ \Delta\rho^2)(1-\frac{\ctw^2}{\stw^2} \Delta\rho +
 \frac{\ctw^4}{\stw^4} \Delta\rho^2),                \nll
\tfflq  &\to& (1+\Delta\alpha(t)+\Delta\alpha(t)^2)(1 + \Delta\rho+ \Delta\rho^2),\nll
\tffql  &\to& (1+\Delta\alpha(t)+\Delta\alpha(t)^2)(1 + \Delta\rho+ \Delta\rho^2),\nll
\tffqq  &\to& (1+\Delta\alpha(t)+\Delta\alpha(t)^2)(1 + \Delta\rho+ \Delta\rho^2)(1+\frac{\ctw^2}{\stw^2}\Delta\rho).
\label{ffllalp0}
\eqa

$\bullet$ {the $G_\mu$ EW scheme}

In the $G_\mu$ EW scheme the coupling constant already contains 
additional $\stw^2$ through the following relation:
\bqa
\alpha_{G_\mu}=\frac{\sqrt{2}G_\mu\mw^2\sw}{\pi}\,,
\eqa
and therefore the replacement for $\alpha_{G_\mu}$ could be written as
\begin{equation}
{\alpha}_{G_\mu} \to \alpha_{G\mu}{\swb}/\sw= 
\alpha_{G_\mu}\left(1+\frac{\cw}{\sw}\Delta\rho\,\right).
\label{FFAexc}
\end{equation}

Since all form factors for $Z$ exchange have an additional $(\sw\cw)^{-1}$
factor, one can write
\bqa
&&\alpha_{G_\mu}\frac{1}{\sw\cw} \to
{\alpha}_{G_\mu}\frac{1}{\sw\cwb }=
\alpha_{G_\mu}\frac{1}{\sw\cw }(1+\Delta\rho+\Delta\rho^2).
\label{FFZexc}
\eqa

Therefore, the form factors at NNLO order read
\bqa
 \tffgg  &\to& 1 + \frac{\ctw^2}{\stw^2}\Delta\rho,         \nll
 \tffll  &\to& 1 + \Delta\rho+ \Delta\rho^2,                \nll
 \tfflq  &\to& (1 + \Delta\rho+ \Delta\rho^2)(1+\frac{\ctw^2}{\stw^2}\Delta\rho),\nll
 \tffql  &\to& (1 + \Delta\rho+ \Delta\rho^2)(1+\frac{\ctw^2}{\stw^2}\Delta\rho),\nll
 \tffqq  &\to& (1 + \Delta\rho+ \Delta\rho^2)(1+\frac{\ctw^2}{\stw^2}\Delta\rho)^2.
\label{ffllgmu}
\eqa

The replacements (\ref{ffllalp0}) and (\ref{ffllgmu}) should be
introduced into helicity amplitudes. After squaring the HAs, all terms
higher than $\Delta\rho^2$,  $\Delta\rho\Delta\alpha$, and $\Delta\alpha^2$ should be omitted. 
To avoid double counting, one should also remove the leading NLO EW contribution (\ref{def-rho-one})
from the terms linear in $\Delta\rho$:
$\Delta\rho \longrightarrow \left( \Delta\rho -\Delta\rhoone\Big|^{\alpha(0)}\right)$ and drop the $\Delta\alpha(t)$ contribution.

We have verified analytically that the results obtained in this way 
agree with the corresponding expressions derived in~\cite{Dittmaier:2009cr}.

\section{Numerical results and comparisons \label{Sect:results}}

In this Section, we present the numerical results for EW RCs to the $\mu^+$ and $\mu^-$
channels of elastic $\mu - e$ 
scattering at the one-loop level obtained by means of the \SANC~ system.

Numerical results are evaluated for two energy points and the following
helicity states of the antimuon/muon ($P_{\mu^{\pm}}$) and electron ($P_{e^-}$)
beam polarization:
\begin{eqnarray}
(P_{\mu^{\pm}},P_{e^-})=
(0,0),(-1,-1),(-1,+1),(+1,-1),(+1,+1)
\label{SetPolarization}
\end{eqnarray}
in the $\alpha(0)$ and $G_{\mu}$ schemes.
Obviously, results for any set of polarizations can be obtained from 
these cross sections.

A comparison of our results for specific contributions at the tree level 
with \CalcHEP{} \cite{Belyaev:2012qa} and
\WHIZARD{}~codes~\cite{Kilian:2007gr}
is given.

We used the following  set of input parameters taken from the PDG (2020)~\cite{ParticleDataGroup:2020ssz}:
\begin{eqnarray}
\begin{array}{ll}
\alpha^{-1}(0) = 137.035999084, & G_{\rm Fermi} = 1.1663787 \cdot 10^{-5}\; \mathrm{GeV}^{-2}\\
\alpha_s(\mz) = 0.1179, & \\
\mz = 91.1876 \; \mathrm{GeV}, & \Gamma_Z = 2.4952 \; \mathrm{GeV}, \\
\mw = 80.379 \; \mathrm{GeV} & \mh = 125.25 \; \mathrm{GeV},\\
m_e = 0.51099895 \; \mathrm{MeV}, & m_\mu = 0.1056583745 \; \mathrm{GeV}, \\
m_\tau = 1.77686 \; \mathrm{GeV}, &\\
m_u = 0.062 \; \mathrm{GeV}, & m_d = 0.083 \; \mathrm{GeV},\\
m_c = 1.5 \; \mathrm{GeV}, & m_s = 0.215 \; \mathrm{GeV},\\
m_t = 172.76 \; \mathrm{GeV}. & m_b = 4.7 \; \mathrm{GeV}.
\nonumber
\end{array}
\end{eqnarray}

We show the results
for the polarized Born, hard bremsstrahlung, one-loop cross sections (pb)
and relative corrections (\%).

\subsection{Comparisons with other codes}

The polarized Born and hard
bremsstrahlung cross sections were cross-checked with the corresponding results
of the \WHIZARD{}~\cite{Kilian:2007gr}
and \CalcHEP{}~\cite{Belyaev:2012qa} programs.

The results are given in the $\alpha(0)$ EW scheme
with fixed $100 \%$ polarized initial states 
for $\sqrt{s}$=250~GeV and
angular cuts $|\cos \vartheta_\mu| \le 0.9$
and $|\cos \vartheta_e| \le 0.9$.
For the hard bremsstrahlung cross sections, an additional cut on the
photon energy 
$E_\gamma \ge \omega = 10^{-4} \sqrt{s}/2$
is applied.

The results of comparisons are shown in Tables~\ref{comphardmup}, \ref{comphardmum}.
Very good agreement within 4-5 digits with the above mentioned codes is found.

\begin{table}[!h]
\centering	
\caption{
The tuned triple comparison between \SANC{}~(the first line), \WHIZARD{}~(the second line), 
and \CalcHEP{}~(the third line) results for the hard bremsstrahlung cross section (pb)
for 100\% polarized $\mu^- e^- \to e^- \mu^- \gamma$ scattering 
and for c.m.s. energy $\sqrt{s} =250$~GeV.
For comparison of the real photon emission we applied an additional
cut on the photon energy
$E_\gamma \ge \omega = 10^{-4} \sqrt{s}/2$.
The angular cuts are $|\cos \vartheta_\mu| \le 0.9$
and $|\cos \vartheta_e| \le 0.9$.
\label{comphardmup}}
\begin{tabular}{lccccc}
	\hline
	\hline
	$P_{\mu^-},P_{e^-}$
	  & 0, 0     &  -1, -1  & -1, 1    & 1, -1    & 1, 1\\
	\hline
	S & 102.42(1) & 157.64(1) & 56.53(1) & 56.52(1) & 139.05(1)\\
	W & 102.43(1) & 157.62(1) & 56.53(1) & 56.54(1) & 139.05(1)\\
	C & 102.43(1) & 157.63(1) & 56.53(1) & 56.52(1) & 139.06(2)\\
	\hline
\end{tabular}
\end{table}

\begin{table}[!h]
\centering	
\caption{
The same as in Table~\ref{comphardmup} but for
$\mu^+ e^- \to e^- \mu^+ \gamma$ scattering.
\label{comphardmum}}
\begin{tabular}{lccccc}
    \hline
    \hline
	$P_{\mu^+},P_{e^-}$
	  & 0, 0     &  -1, -1  & -1, 1    & 1, -1     & 1, 1\\
	\hline
	S & 91.63(1) & 77.27(1) & 99.88(1) & 112.12(1) & 77.28(1)\\
	W & 91.63(1) & 77.28(1) & 99.90(1) & 112.12(1) & 77.28(1)\\
	C & 91.63(1) & 77.27(1) & 99.89(1) & 112.11(1) & 77.27(1)\\
    \hline
\end{tabular}
\end{table}

\subsection{One-loop cross sections and relative corrections}

\subsubsection{The case of $\sqrt{s}$=250 GeV c.m.s. energy}

In Tables~\ref{Table:delta_250plus}--\ref{Table:delta_250minus}
we present the values of the Born cross sections (in pb)
as well as the relative corrections 
$$\delta^{\rm Y} = \frac{\sigma^{\rm Y}}{\sigma^{\rm Born}},\, \%$$
for reactions (\ref{muplus}--\ref{muminus})
for the c.m.s. energy $\sqrt{s}=250$~GeV with angular
cuts $|\cos \vartheta_\mu| \le 0.9$
and $|\cos \vartheta_e| \le 0.9$.
We consider the unpolarized and fully (100\%) polarized
initial beams (\ref{SetPolarization}) in the 
$\alpha(0)$ and $G_\mu$ EW schemes.

In the tables, particular contributions of the NLO EW relative corrections
as well as the leading h.o. corrections are shown. We split the
complete one-loop contribution into two gauge-invariant subsets of the 
diagrams $\delta^{\rm QED}$ and  $\delta^{\rm weak}$, where
$\delta^{\rm weak}$ includes vacuum polarization (vp) as well as pure weak-interaction 
contributions; $\delta^{\rm weak - vp} = \delta^{\rm weak} - \delta^{\rm vp}$.

\begin{table}[ht]
\begin{center}
\caption{Integrated Born cross sections and
relative corrections
for $\mu^+ e^- \to e^- \mu^+ (\gamma)$ scattering
for the c.m.s. energy $\sqrt{s}=250$~GeV
and the set (\protect\ref{SetPolarization}) of polarization degrees
of the initial particles in the $\alpha(0)$ and $G_\mu$ EW schemes.}
\label{Table:delta_250plus}
\begin{tabular}{llllll}
\hline\hline
$P_{\mu^+}$, $P_{e^-}$  & $0,0$   & $-1,-1$    & $-1,+1$    & $+1,-1$  & $+1,+1$
\\ \hline
\multicolumn{6}{c}{$\alpha(0)$ EW scheme} \\
$\sigma^{\text{Born}}$, pb
                     & 66.487(1)  & 55.333(1)  & 73.186(1)  & 82.097(1) & 55.333(1)
\\
{$\delta^{\text{QED}}, \%$}
                     & -1.936(1)  & -0.481(1)  & -2.933(1)   & -3.013(1)  & -0.482(1)
\\
{$\delta^{\text{VP}}$, $\%$}
                     & 11.466(1)    & 13.729(2)   & 10.151(1)  & 9.586(1)   & 13.729(2)
\\
{$\delta^{\text{weak - VP}}$, $\%$}
                     & -0.396(1)    & -1.758(1)   & 2.297(1)   & -0.962(1)   & -1.758(1)
\\
{$\delta^{\text{ho}}, \%$}
                     & 1.032(1) & 0.929(1) & 0.895(1) & 1.295(1) & 0.929(1)
\\
\multicolumn{6}{c}{$G_\mu$ EW scheme} \\
$\sigma^{\text{Born}}$, pb
                     & 71.458(1)  & 59.470(1)  & 78.658(1)  & 88.234(1) & 59.470(1)
\\
{$\delta^{\text{QED}}, \%$}
                     & -1.935(2)  & -0.481(2)  & -2.930(2)   & -3.007(2)  & -0.482(2)
\\
{$\delta^{\text{VP}}$, $\%$}
                     & 5.568(1)   & 6.705(1)   & 4.899(1)  & 4.630(1)   & 6.705(2)
\\
{$\delta^{\text{weak - VP}}$, $\%$}
                     & -0.391(1)  & -0.626(1)  & 1.656(1)  & -1.891(1)  & -0.626(1)
\\
{$\delta^{\text{ho}}, \%$}
                     & -0.456(1)   & -0.512(1)   & -0.520(1)   & -0.322(1)  & -0.512(1)
\\ \hline\hline
\end{tabular}
\end{center}
\end{table}

\begin{table}[ht]
\begin{center}
\caption{The same as in Table~\ref{Table:delta_250plus}
but for $\mu^- e^- \to e^- \mu^- (\gamma$) scattering.
}
\label{Table:delta_250minus}
\begin{tabular}{llllll}
\hline\hline
$P_{\mu^+}$, $P_{e^-}$ & $0,0$   & $-1,-1$    & $-1,+1$   & $+1,-1$   & $+1,+1$
\\ \hline
\multicolumn{6}{c}{$\alpha(0)$ EW scheme} \\
$\sigma^{\text{Born}}$, pb
                     & 75.231(1) & 115.076(1) & 42.157(1) & 42.157(1) & 101.538(1)
\\
{$\delta^{\text{QED}}, \%$}
                     & -2.085(1)  & -1.912(1)  & -2.682(1)  & -2.683(1) & -1.781(1)
\\
{$\delta^{\text{VP}}, \%$}
                     & 10.849(1)   & 9.602(1) & 13.305(1)  & 13.305(1) & 10.220(1)    
\\
{$\delta^{\text{weak - VP}}, \%$}
                     & -0.161(1)   & -1.476(1) & -1.540(1)  & -1.540(1) & 2.474(1)
\\
{$\delta^{\text{ho}}, \%$}
                     & 1.089(1) & 1.365(1) & 0.907(1) & 0.907(1) & 0.926(1)
\\
\multicolumn{6}{c}{$G_\mu$ EW scheme} \\
$\sigma^{\text{Born}}$, pb
                     & 80.855(1) & 123.679(1) & 45.309(1) & 45.309(1) & 109.128(1)
\\
{$\delta^{\text{QED}}, \%$}
                     & -2.082(2) & -1.911(1)  & -2.685(2)  & -2.685(2)  & -1.780(1)
\\
{$\delta^{\text{VP}}, \%$}
                     & 5.295(1)   & 4.729(1) & 6.393(1)  & 6.393(1) & 5.027(1)
\\
{$\delta^{\text{weak - VP}}, \%$}
                     & -0.501(1) & -2.495(1) & -0.519(1) & -0.519(1) & 1.775(1)
\\
{$\delta^{\text{ho}}, \%$}
                     & -0.436(1) & -0.306(1) & -0.511(1) & -0.511(1) & -0.522(1)
\\ \hline\hline 
\end{tabular}
\end{center}
\end{table}

It is convenient to discuss both tables in the same manner.
A comparison of the cross sections with different values of polarization
has demonstrated the significance of polarization effects.
Namely, for the $\mu^+$ channel the cross section $\sigma_{+-}$
is about 1.5 times larger than $\sigma_{--,++}$ and
about 1.2 times larger than the unpolarized cross section.
As for the $\mu^-$ channel, the cross section $\sigma_{--}$
is about 2.5 times larger than $\sigma_{-+,+-}$ and
about 1.5 times larger than the unpolarized cross section.

The QED part of the relative RCs for the unpolarized and fully polarized cases 
in the $\alpha(0)$ and $G_\mu$ EW schemes is negative 
and has a maximum of about 3\% in the $\mu^+$ channel for $\delta^{\rm QED}_{+-,-+}$
(the same situation is in the $\mu^-$ channel, i.e., the maximum is about 2.7\% 
for the same polarization values),
and has a minimum of about -0.5\% in the $\mu^+$ channel for $\delta^{\rm QED}_{--,++}$ 
(about (1.8-1.9)\% in the $\mu^-$ channel).

The main contribution to the weak correction is due to the vacuum polarization
$\delta^{\rm vp}$
which is positive and gives about 10-13\% in the $\alpha(0)$ EW scheme 
(5-7\% in the $G_\mu$ scheme) for both $\mu^+$ and $\mu^-$ channels.

The rest of the weak correction $\delta^{\rm weak-vp}$ is also negative
and it can reach up to 1.5-2\% and even can change the sign.

The leading higher-order corrections are positive and equal to about 1\%
in the $\alpha(0)$ EW scheme while in the $G_\mu$ one they are negative and
equal to about half a per cent.

All weak and h.o. corrections strongly depend on the choice of the EW scheme,
and the total weak corrections in $G_\mu$ scheme are smaller by about
5-6\% than in the $\alpha(0)$ one.

\begin{table}[ht]
\begin{center}
\caption{Weak
and higher-order corrected cross sections 
for $\mu^+ e^- \to e^- \mu^+ (\gamma$) scattering
for the c.m.s. energy $\sqrt{s}=250$~GeV
and the set (\protect\ref{SetPolarization}) of polarization degrees
of the initial particles in the $\alpha(0)$ and $G_\mu$ EW schemes.}
\label{Table:weakho_250plus}
\begin{tabular}{llllll}
\hline\hline
$P_{\mu^+}$, $P_{e^-}$  & $0,0$   & $-1,-1$    & $-1,+1$    & $+1,-1$  & $+1,+1$
\\ \hline
{$\sigma^{\text{weak}}_{\alpha(0)}$, pb}
                     & 73.846(1)  & 61.956(1) & 82.295(1)  & 89.175(1) & 61.956(1)
\\
{$\sigma^{\text{weak}}_{G_\mu}$, pb}
                     & 75.156(1)  & 63.084(1)  & 83.812(1)  & 90.642(1) & 63.084(1)
\\
{$\delta^{\text{weak}}_{G_\mu / \alpha(0)}, \%$}
                     & 1.77       & 1.82       & 1.84       & 1.65      & 1.82
\\
{$\sigma^{\text{weak+ho}}_{\alpha(0)}$, pb}
                     & 74.533(1)  & 62.471(1) & 82.951(1)  & 90.240(1)  & 62.471(1)
\\
{$\sigma^{\text{weak+ho}}_{G_\mu}$, pb}
                     & 74.830(1)  & 62.779(1)  & 83.405(1)  & 90.359(1) & 62.779(1)
\\
{$\delta^{\text{weak+ho}}_{G_\mu / \alpha(0)}, \%$}
                     & 0.40       & 0.50       & 0.55      & 0.13     & 0.50
\\ \hline\hline 
\end{tabular}
\end{center}
\end{table}

\begin{table}[h!]
\begin{center}
\caption{
The same as in Table~\ref{Table:weakho_250plus}
but for $\mu^- e^- \to e^- \mu^- (\gamma$) scattering.
}
\label{Table:weakho_250min}
\begin{tabular}{llllll}
\hline\hline
$P_{\mu^+}$, $P_{e^-}$  & $0,0$   & $-1,-1$    & $-1,+1$    & $+1,-1$  & $+1,+1$
\\ \hline
{$\sigma^{\text{weak}}_{\alpha(0)}$, pb}
                     & 82.272(1)  & 124.427(1) & 47.117(1) & 47.117(1) & 114.427(1)
\\
{$\sigma^{\text{weak}}_{G_\mu}$, pb}
                     & 84.732(1)  & 126.441(1) & 47.969(1) & 47.969(1) & 116.551(1)
\\
{$\delta^{\text{weak}}_{G_\mu / \alpha(0)}, \%$}
                     & 2.99       &  1.62      & 1.81       &  1.81    & 1.86
\\
{$\sigma^{\text{weak+ho}}_{\alpha(0)}$, pb}
                     & 84.091(1)  & 125.999(1) & 47.499(1) & 47.499(1) & 115.368(1)
\\
{$\sigma^{\text{weak+ho}}_{G_\mu}$, pb}
                     & 84.379(1)  & 126.062(1) & 47.738(1) & 47.738(1)  & 115.981(1)
\\
{$\delta^{\text{weak+ho}}_{G_\mu / \alpha(0)}, \%$}
                     & 0.34       & 0.05       & 0.05       & 0.50     & 0.53
\\ \hline\hline 
\end{tabular}
\end{center}
\end{table}

In Tables~\ref{Table:weakho_250plus} and \ref{Table:weakho_250min}, the 
integrated cross sections for the weak and leading higher-order corrections are shown
for both $\mu^+$ and $\mu^-$ channels. The results are calculated in the
$\alpha(0)$ and $G_\mu$ schemes, their difference
\begin{eqnarray}
\delta_{G_\mu/\alpha(0)} = \frac{\sigma_{G_\mu}}{\sigma_{\alpha(0)}} - 1,\, \%
\label{rgmual0}
\end{eqnarray}
is also shown. Ratio~(\ref{rgmual0}) 
shows the stabilization of the results and can be considered as
an estimation of the theoretical uncertainty of the weak and h.o. contributions.
As it is well known, the difference between two EW schemes in the LO is just the
ratio of the EW couplings and gives about $\delta^{\rm LO}_{G_\mu/\alpha(0)} = 7\%$. 
As it is seen from the tables, the weak contribution reduces the difference almost by
one half to about $\delta^{\rm weak}_{G_\mu/\alpha(0)} = 2-3\%$. Even more, the
sum of the weak and h.o. contributions reduces the difference almost by
one half or one third to about $\delta^{\rm weak+h.o.}_{G_\mu/\alpha(0)} = 0.05-0.5\%$
depending on the value of polarization degrees and the reaction channel. We consider 0.5\%
as a rather big difference which is caused by using the Mandelstam variable $t$ as
a scale of the $\Delta\alpha(t)$ (\ref{alphat}) quantity in calculations of the 
h.o. contributions within the $\alpha(0)$ EW scheme.

\begin{figure}
    \centering
    \includegraphics[]{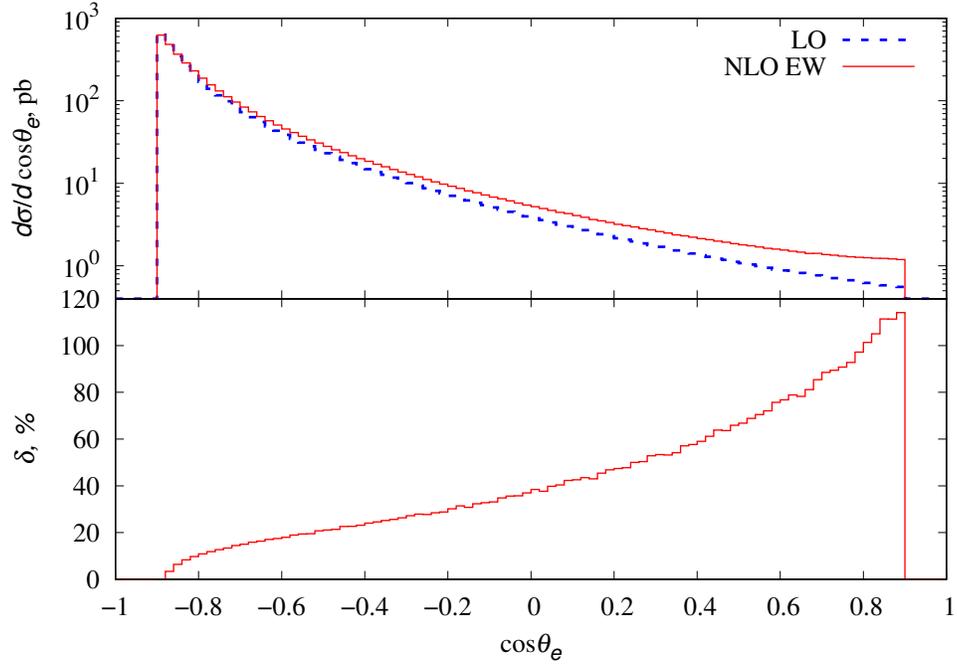}
    \caption{The LO and NLO EW cross sections (upper panel) and relative corrections (lower panel)
    of the $\mu^+ e^- \to e^- \mu^+(\gamma)$ process
    for the c.m.s. energy $\sqrt{s}=250$~GeV
    as a function of $\cos \theta_e$.}
    \label{fig:mup-cosel}
\end{figure}

\begin{figure}
    \centering
    \includegraphics{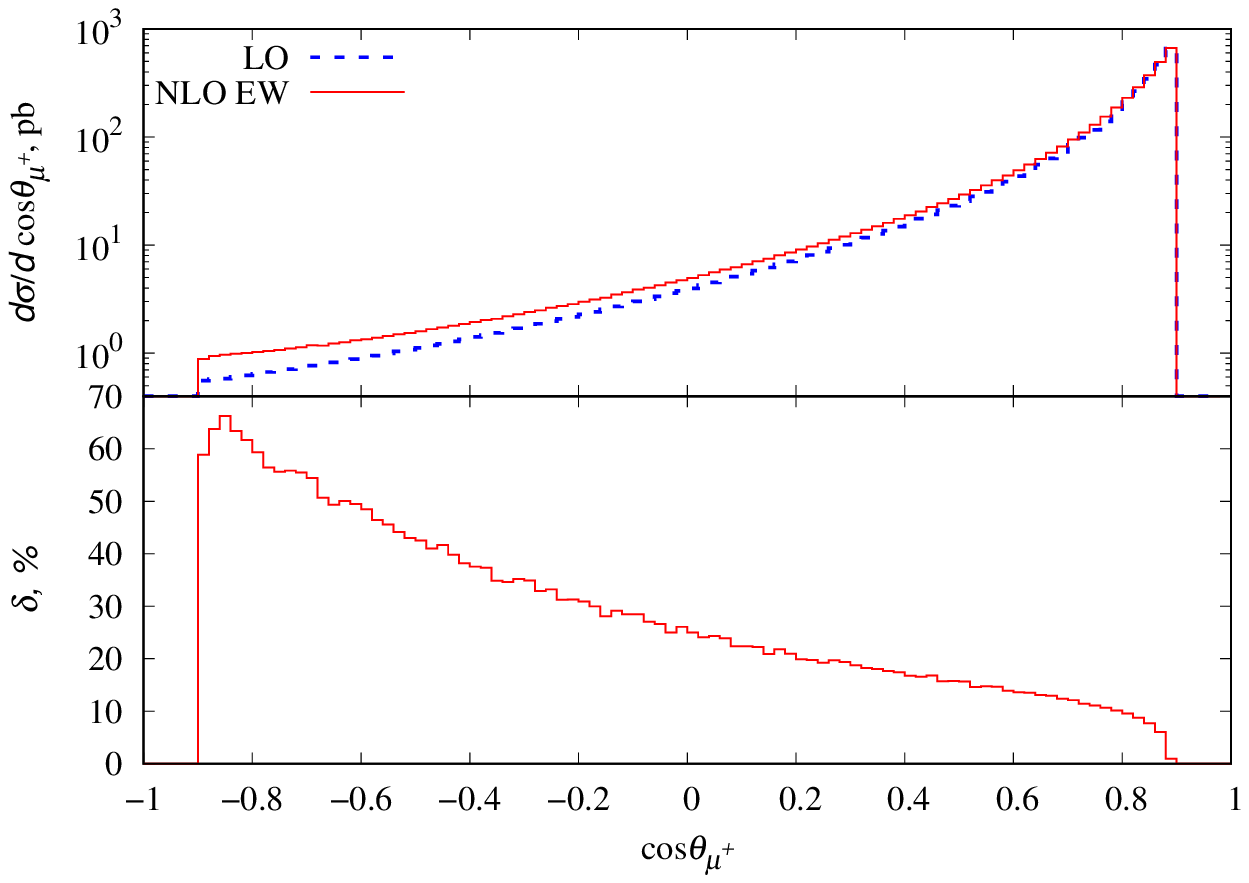}
    \caption{The same as in figure~\ref{fig:mup-cosel} but for $\cos \theta_{\mu^+}$.}
    \label{fig:mup-cosmu}
\end{figure}

\begin{figure}
    \centering
    \includegraphics[]{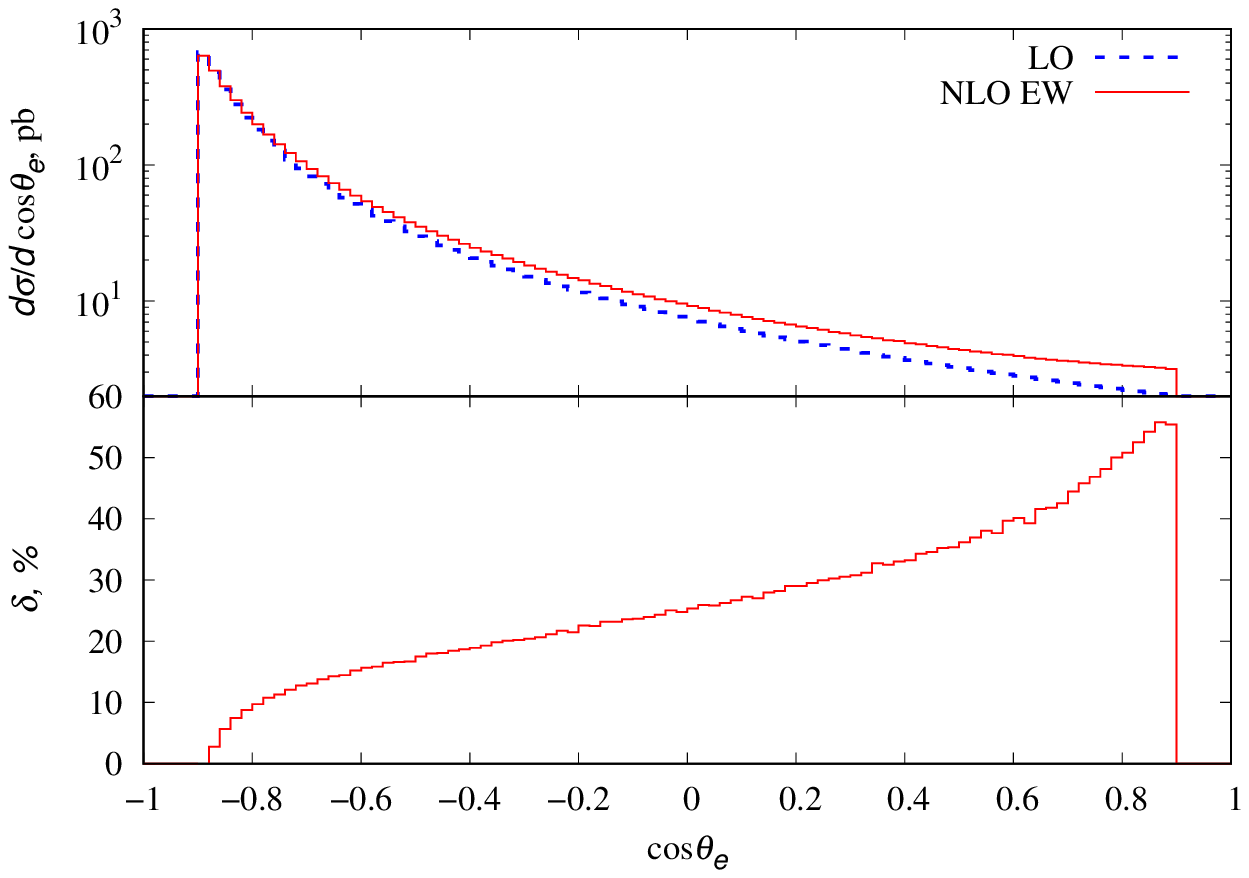}
    \caption{The LO and NLO EW cross sections (upper panel) and relative corrections (lower panel)
    of the $\mu^- e^- \to e^- \mu^-(\gamma)$ process
    for the c.m.s. energy $\sqrt{s}=250$~GeV
    as a function of $\cos \theta_e$.}
    \label{fig:mum-cosel}
\end{figure}

\begin{figure}
    \centering
    \includegraphics{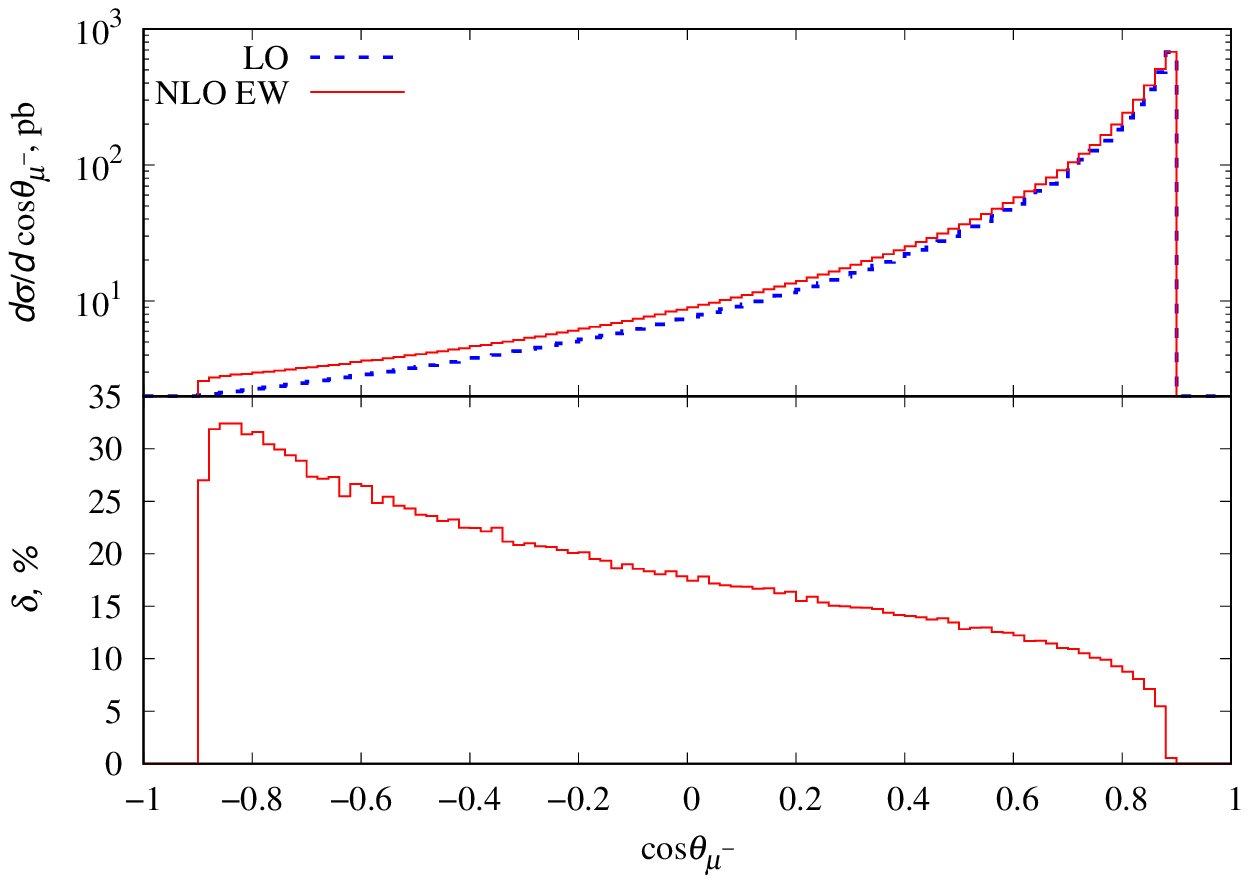}
    \caption{The same as in figure~\ref{fig:mum-cosel} but for $\cos \theta_{\mu^-}$.}
    \label{fig:mum-cosmu}
\end{figure}

Figures~\ref{fig:mup-cosel},\ref{fig:mum-cosel} and~\ref{fig:mup-cosmu},\ref{fig:mum-cosmu} 
show the differential 
cross sections for LO and NLO EW and relative corrections
as functions of the $\cos \theta_e$ and $\cos \theta_\mu^\pm$, respectively.
The maxima for the LO and NLO EW cross section are at the $t \to 0$ ($\cos{\theta_e} \simeq -1$
of the final electron and $\cos{\theta_\mu^\pm} \simeq 1$ of the final (anti-)muon). 
At the same time the relative corrections have maxima at opposite
values of angles where real photon contribution dominates.\footnote{It should be noted that in~\cite{Bardin:2017mdd} all figures are plotted 
as a function of the $\cos({\widehat{ {\vec p_2}{\vec p_3}}})$.} 

\subsubsection{The case of the laboratory system energy $E_\mu$=150 GeV, $E_e=m_e$}

In order to validate the code, we have calculated NLO EW and h.o. corrections for the setup
of the MUonE experiment proposed in~\cite{Alacevich:2018vez}. Namely, we have used Setup~1:
$E_\mu$=150~GeV, $E_e=m_e$ ($\sqrt{s} \simeq$ 0.405541~GeV), $\theta_e, \theta_\mu \le$ 100~mrad, 
$E_e \ge 0.2$~GeV.

The results are shown in Tables~\ref{Table:delta_04plus} and~\ref{Table:delta_04minus}
for both muon channels and for the set of polarization degrees~(\ref{SetPolarization})
of the initial particles in the $\alpha(0)$ EW scheme. In the tables we show the integrated 
Born cross sections and the relative corrections for the QED part, the vacuum polarization part 
of the weak contribution, and the part of h.o. corrections proportional to $\Delta\alpha(t)^2$.
The rest of the weak contribution $\delta^{\rm weak-vp}$ which is about $10^{-4} \%$ and
the rest of h.o. corrections (not proportional to $\Delta\alpha(t)^2$) is about
$10^{-8} \%$, they are omitted in the tables.

Tables~\ref{Table:delta_04plus} and~\ref{Table:delta_04minus} show that the effect of 
the initial particle polarization
changes the third digits in $\delta^{\rm QED}$ and $\delta^{\rm vp}$.
The integrated cross section $\sigma_{--,++}$
is larger than $\sigma_{unp}$ by about 0.8\% for both $\mu^+$ and $\mu^-$ channels.
The h.o. contribution from $\Delta\alpha(t)^2$ is about $6\cdot10^{-3} \%$.

\begin{table}[ht]
\begin{center}
\caption{Integrated Born cross section
and relative corrections
for $\mu^+ e^- \to e^- \mu^+ (\gamma$) scattering
for the laboratory system energy $E_\mu$=150~GeV
and the set (\protect\ref{SetPolarization}) of the initial particle 
polarization degrees in the $\alpha(0)$ EW scheme.}
\label{Table:delta_04plus}
\begin{tabular}{llllll}
\hline\hline
$P_{\mu^+}$, $P_{e^-}$  & $0,0$   & $-1,-1$   & $-1,+1$   & $+1,-1$  & $+1,+1$
\\ \hline
$\sigma^{\text{Born}}$, $\mu$b
                     & 1265.1(1)  & 1275.3(1) & 1254.8(1) & 1254.8(1) & 1275.3(1)
\\
$\delta^{\text{QED}}$, $\%$
                     & 4.762(1) & 4.766(1)   & 4.757(1)   & 4.759(1)   & 4.765(1)
\\
{$\delta^{\text{VP}}$, $\%$}
                     & 0.940(1)    & 0.943(1)   & 0.936(1)   & 0.936(1)   & 0.943(1)
\\
{$\delta^{\text{ho}} (\Delta\alpha^2)$, $\%$}
                     & 0.006(1)    & 0.006(1)   & 0.006(1)   & 0.006(1)   & 0.006(1)
\\ \hline\hline 
\end{tabular}
\end{center}
\end{table}

\begin{table}[ht]
\begin{center}
\caption{
The same as in Table~\ref{Table:delta_04plus}
but for $\mu^- e^- \to e^- \mu^- (\gamma$) scattering.
}
\label{Table:delta_04minus}
\begin{tabular}{llllll}
\hline\hline
$P_{\mu^-}$, $P_{e^-}$  & $0,0$      & $-1,-1$    & $-1,+1$    & $+1,-1$  & $+1,+1$
\\ \hline
$\sigma^{\text{Born}}$, $\mu$b
                     & 1265.1(1) & 1275.3(1) & 1254.8(1) & 1254.8(1) & 1275.3(1)
\\
{$\delta^{\text{QED}}, \%$}
                     & 4.624(1)   & 4.608(1)   & 4.640(1)   & 4.639(1)   & 4.608(1)
\\
{$\delta^{\text{VP}}, \%$}
                     & 0.940(1)   & 0.940(1)   & 0.940(1)   & 0.940(1)   & 0.940(1)
\\
{$\delta^{\text{ho}}(\Delta\alpha^2$), $\%$}
                     & 0.006(1)    & 0.006(1)   & 0.006(1)   & 0.006(1)   & 0.006(1)
\\ \hline\hline 
\end{tabular}
\end{center}
\end{table}

\begin{figure}[ht]
    \centering
    \includegraphics[]{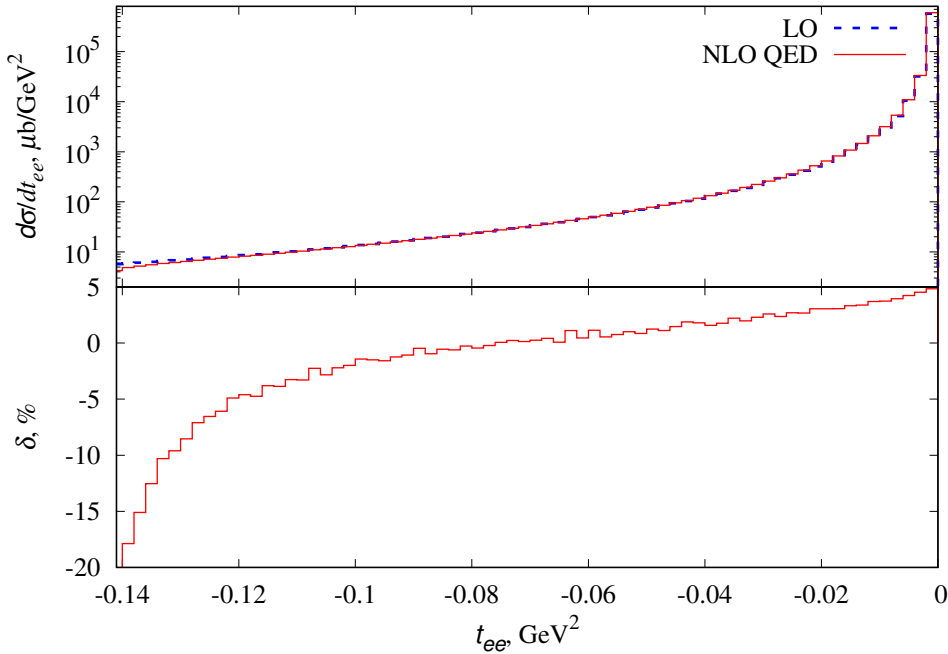}
    \caption{The LO and NLO QED cross sections (upper panel) and 
    relative corrections (lower panel) of the $\mu^+ e^- \to e^- \mu^+(\gamma)$ process
    for the laboratory system energy $E_\mu$=150~GeV
    as a function of $t_{ee}=(p_2-p_3)^2$.}
    \label{fig:tee}
\end{figure}

\begin{figure}[ht]
    \centering
    \includegraphics{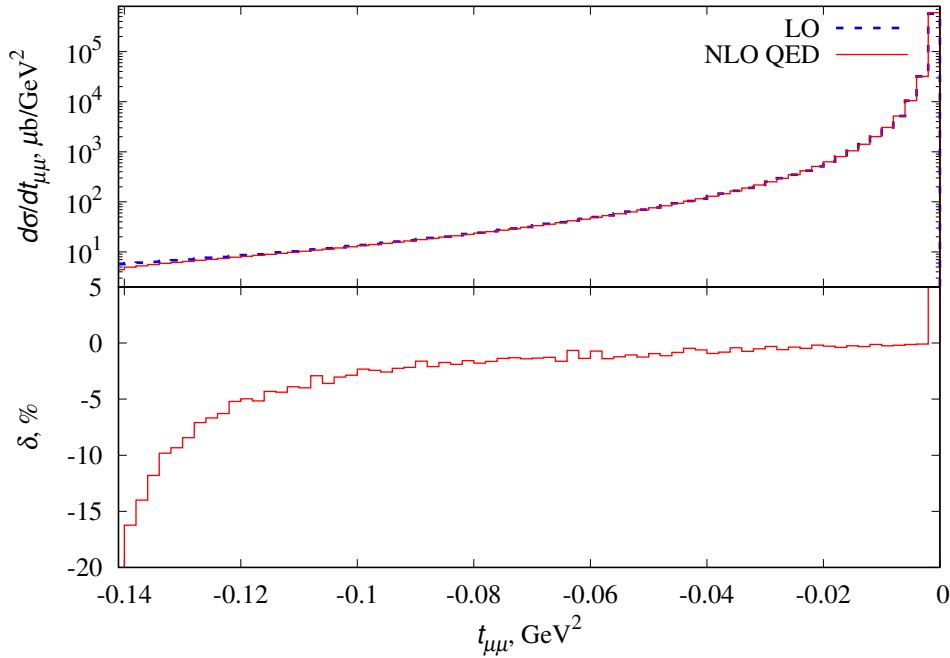}
    \caption{The same as in figure~\ref{fig:tee} but for $t_{\mu\mu}=(p_1-p_4)^2$.}
    \label{fig:tmm}
\end{figure}

\begin{figure}[ht]
    \centering
    \includegraphics{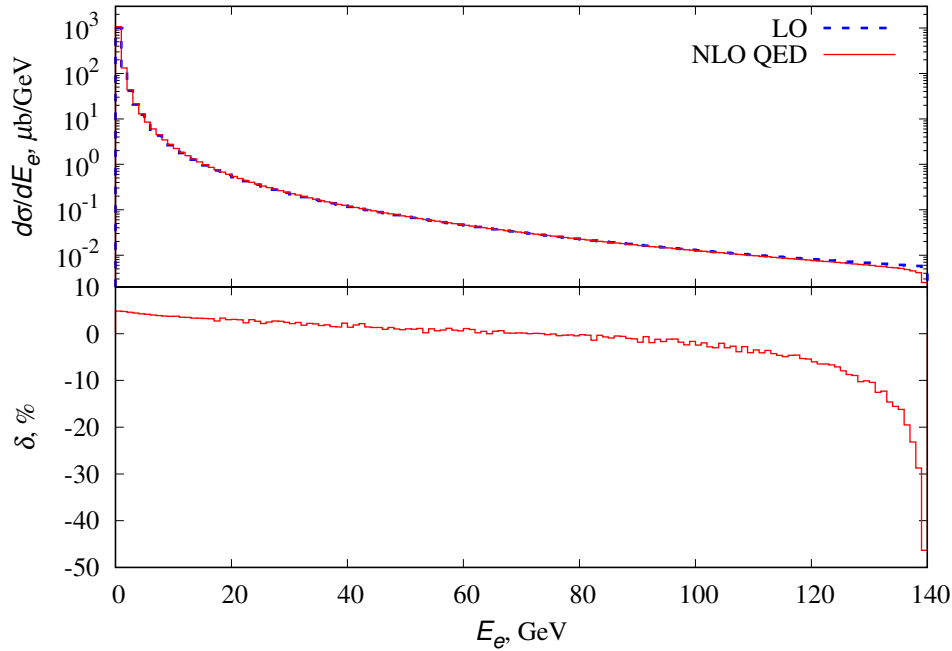}
    \caption{The same as in figure~\ref{fig:tee} but for electron energy $E_e$.}
    \label{fig:eel}
\end{figure}

\begin{figure}[ht]
    \centering
    \includegraphics{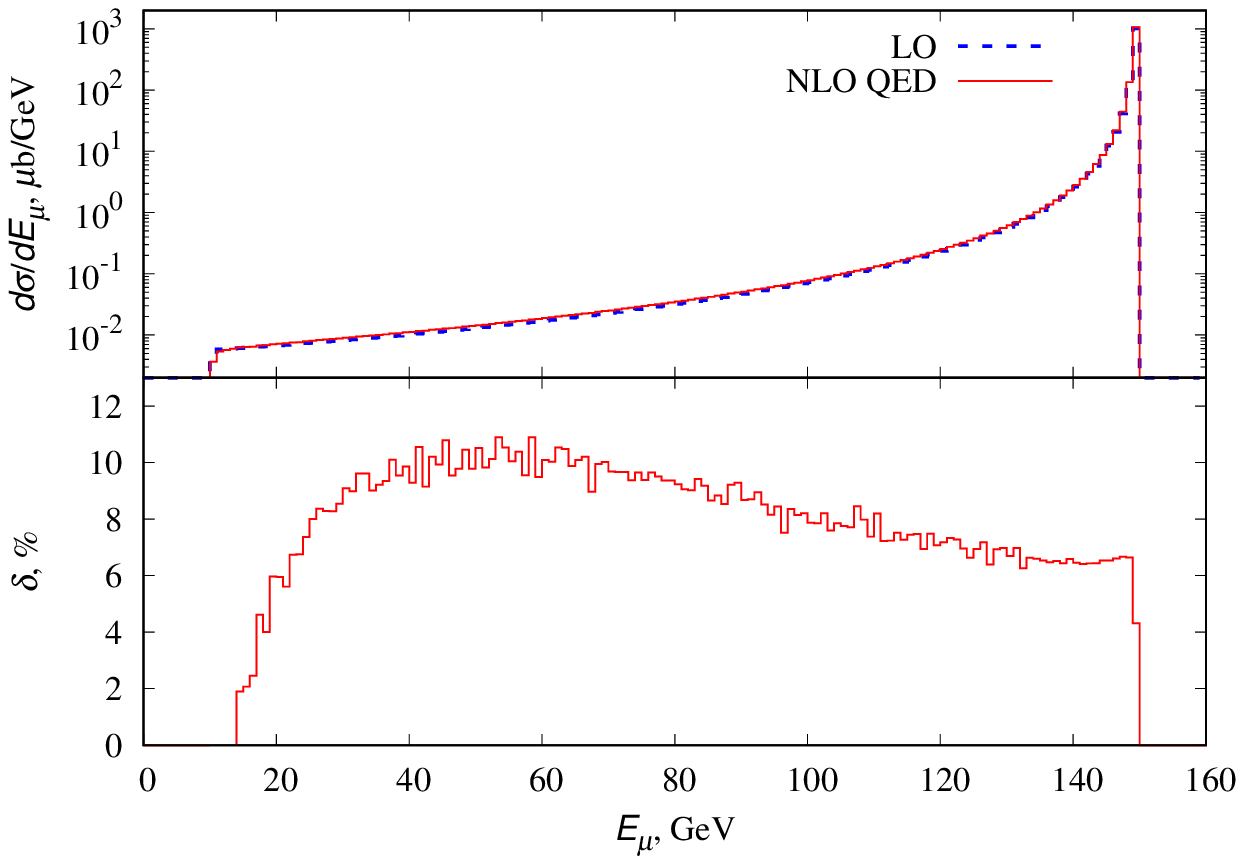}
    \caption{The same as in figure~\ref{fig:tee} but for muon energy $E_\mu$.}
    \label{fig:emu}
\end{figure}



Figures~\ref{fig:tee} and~\ref{fig:tmm} show the differential cross
for LO and NLO QED and relative corrections
as functions of the variables  $t_{ee}=(p_2-p_3)^2$ and $t_{\mu\mu}=(p_1-p_4)^2$, respectively.
In figures~\ref{fig:eel} and~\ref{fig:emu} the distributions 
on the final electron $E_e$ 
and muon $E_\mu$ energies are presented.
All distributions look very similar to those given in~\cite{Alacevich:2018vez}.

\section{Conclusions \label{Sect:Concl}}

In this paper, we have described the implementation 
of complete one-loop EW corrections to elastic
$\mu - e$ scattering within the {\SANC}~system framework.
The relevant contributions to the cross section are calculated analytically
using the helicity amplitude approach, which allows to take into account
any polarization. 
The helicity amplitudes were used for
Born and virtual parts as well as for the soft and hard photon bremsstrahlung,
taking into account the masses of the initial and final fermions.

The numerical results are also presented.
The effects of initial beam polarization 
were analyzed for fully polarized states. 
Two energy points were considered: the c.m.s. energy $\sqrt{s}=250$~GeV 
and the laboratory system energy $E_\mu$=150~GeV. 

The calculated polarized cross sections at the tree level
for Born and hard photon brems\-strahlung
were thoroughly compared with the results of {\tt CalcHEP} and {\tt WHIZARD}, and
very good agreement with the above-mentioned codes
was observed.

It was found that cross sections strongly depend on polarization degrees of initial beams. 
As a result, the polarization effects are significant and 
give increase in the cross sections at definite polarization degrees.

The complete one-loop as well as the leading higher-order
corrections were analyzed.
Higher-order terms were introduced by using
the parameters $\Delta\alpha$ and $\Delta\rho$.

For the c.m.s. energy $\sqrt{s}=250$~GeV  
calculations in $\alpha(0)$ and $G_\mu$ EW schemes were performed.
The sum of weak and higher-order contributions
reduces the difference between the results in two EW schemes
to about $0.05-0.5\%$
depending on the polarization values and the reaction channel.
This part of our study can be considered as a preliminary step
in preparation of a contribution to the physical program of the
proposed $\mu$TRISTAN experiment~\cite{Hamada:2022mua}.

For the laboratory system energy $E_\mu$=150 GeV, we present cross sections and distributions calculated with up to date input parameters.

The forthcoming part of our work on this process will be devoted to calculation 
of leading and next-to-leading large logarithmic corrections 
in higher orders $\alpha^n L^n$ and $\alpha^n L^{n-1}$, where $L=\ln(Q^2/m_e^2)$
and $n\geq 3$.

\section{Acknowledgments}
This research was funded by RFBR grant 20-02-00441.


\begingroup\begin{thebibliography}{10}

\bibitem{Kukhto:1987uj}
T.~V. Kukhto, N.~M. Shumeiko, and S.~I. Timoshin, {\em J. Phys. G} {\bf 13}
  (1987) 725--734.

\bibitem{Kaiser:2010zz}
N.~Kaiser, {\em J. Phys. G} {\bf 37} (2010) 115005.

\bibitem{SpinMuonSMC:1997ixm}
{Spin Muon (SMC)} Collaboration, D.~Adams {\em et al.}, {\em Phys. Lett. B}
  {\bf 396} (1997) 338--348.

\bibitem{Bardin:1997nc}
D.~Y. Bardin and L.~Kalinovskaya,
  \href{http://www.arXiv.org/abs/hep-ph/9712310}{{\tt hep-ph/9712310}}.

\bibitem{Hamada:2022mua}
Y.~Hamada, R.~Kitano, R.~Matsudo, H.~Takaura, and M.~Yoshida,
  \href{http://www.arXiv.org/abs/2201.06664}{{\tt 2201.06664}}.

\bibitem{Abbiendi:2016xup}
G.~Abbiendi {\em et al.}, {\em Eur. Phys. J. C} {\bf 77} (2017), no.~3 139,
  \href{http://www.arXiv.org/abs/1609.08987}{{\tt 1609.08987}}.

\bibitem{Abbiendi:2677471}
G.~Abbiendi, ``{Letter of Intent: the MUonE project}'', preprint CERN, Geneva
  (6, 2019).

\bibitem{Masiero:2020vxk}
A.~Masiero, P.~Paradisi, and M.~Passera, {\em Phys. Rev. D} {\bf 102} (2020),
  no.~7 075013, \href{http://www.arXiv.org/abs/2002.05418}{{\tt 2002.05418}}.

\bibitem{Dev:2020drf}
P.~S.~B. Dev, W.~Rodejohann, X.-J. Xu, and Y.~Zhang, {\em JHEP} {\bf 05} (2020)
  053, \href{http://www.arXiv.org/abs/2002.04822}{{\tt 2002.04822}}.

\bibitem{Alacevich:2018vez}
M.~Alacevich, C.~M. Carloni~Calame, M.~Chiesa, G.~Montagna, O.~Nicrosini, and
  F.~Piccinini, {\em JHEP} {\bf 02} (2019) 155,
  \href{http://www.arXiv.org/abs/1811.06743}{{\tt 1811.06743}}.

\bibitem{CarloniCalame:2019mbo}
C.~M. Carloni~Calame, M.~Chiesa, G.~Montagna, O.~Nicrosini, and F.~Piccinini,
  {\em EPJ Web Conf.} {\bf 212} (2019) 05002.

\bibitem{CarloniCalame:2020yoz}
C.~M. Carloni~Calame, M.~Chiesa, S.~M. Hasan, G.~Montagna, O.~Nicrosini, and
  F.~Piccinini, {\em JHEP} {\bf 11} (2020) 028,
  \href{http://www.arXiv.org/abs/2007.01586}{{\tt 2007.01586}}.

\bibitem{Budassi:2021twh}
E.~Budassi, C.~M. Carloni~Calame, M.~Chiesa, C.~L. Del~Pio, S.~M. Hasan,
  G.~Montagna, O.~Nicrosini, and F.~Piccinini, {\em JHEP} {\bf 11} (2021) 098,
  \href{http://www.arXiv.org/abs/2109.14606}{{\tt 2109.14606}}.

\bibitem{Banerjee:2020rww}
P.~Banerjee, T.~Engel, A.~Signer, and Y.~Ulrich, {\em SciPost Phys.} {\bf 9}
  (2020) 027, \href{http://www.arXiv.org/abs/2007.01654}{{\tt 2007.01654}}.

\bibitem{Ulrich:2020frs}
Y.~Ulrich, ``{McMule -- QED Corrections for Low-Energy Experiments}'', other
  thesis, 8, 2020.

\bibitem{Fael:2019nsf}
M.~Fael and M.~Passera, {\em Phys. Rev. Lett.} {\bf 122} (2019), no.~19 192001,
  \href{http://www.arXiv.org/abs/1901.03106}{{\tt 1901.03106}}.

\bibitem{Balzani:2020yxg}
E.~Balzani, ``{Leptonic QED Contributions to Muon-Electron Scattering at
  NNLO}'', Master's thesis, Padua U., 2020.

\bibitem{Banerjee:2020tdt}
P.~Banerjee {\em et al.}, {\em Eur. Phys. J. C} {\bf 80} (2020), no.~6 591,
  \href{http://www.arXiv.org/abs/2004.13663}{{\tt 2004.13663}}.

\bibitem{Andonov:2004hi}
A.~Andonov, A.~Arbuzov, D.~Bardin, S.~Bondarenko, P.~Christova,
  L.~Kalinovskaya, G.~Nanava, and W.~von Schlippe, {\em Comput. Phys. Commun.}
  {\bf 174} (2006) 481--517, [Erratum: Comput.Phys.Commun. 177, 623--624
  (2007)], \href{http://www.arXiv.org/abs/hep-ph/0411186}{{\tt
  hep-ph/0411186}}.

\bibitem{Bardin:2019zsp}
D.~Y. Bardin {\em et al.}, {\em Phys. Part. Nucl.} {\bf 50} (2019), no.~4
  395--432.

\bibitem{Bardin:2017mdd}
D.~Bardin, Y.~Dydyshka, L.~Kalinovskaya, L.~Rumyantsev, A.~Arbuzov, R.~Sadykov,
  and S.~Bondarenko, {\em Phys. Rev. D} {\bf 98} (2018), no.~1 013001,
  \href{http://www.arXiv.org/abs/1801.00125}{{\tt 1801.00125}}.

\bibitem{Bondarenko:2020hhn}
S.~Bondarenko, Y.~Dydyshka, L.~Kalinovskaya, R.~Sadykov, and V.~Yermolchyk,
  {\em Phys. Rev. D} {\bf 102} (2020), no.~3 033004,
  \href{http://www.arXiv.org/abs/2005.04748}{{\tt 2005.04748}}.

\bibitem{Andonov:2002xc}
A.~Andonov, D.~Bardin, S.~Bondarenko, P.~Christova, L.~Kalinovskaya, and
  G.~Nanava, {\em Phys. Part. Nucl.} {\bf 34} (2003) 577--618, [Fiz. Elem.
  Chast. Atom. Yadra 34,1125(2003)],
\href{http://www.arXiv.org/abs/hep-ph/0207156}{{\tt hep-ph/0207156}}.

\bibitem{Kilian:2007gr}
W.~Kilian, T.~Ohl, and J.~Reuter, {\em Eur. Phys. J.} {\bf C71} (2011) 1742,
\href{http://www.arXiv.org/abs/0708.4233}{{\tt 0708.4233}}.

\bibitem{Passarino:1978jh}
G.-P. Passarino and M.~J.~G. Veltman, {\em Nucl. Phys. B} {\bf 160} (1979)
  151--207.

\bibitem{MoortgatPick:2005cw}
G.~Moortgat-Pick {\em et al.}, {\em Phys. Rept.} {\bf 460} (2008) 131--243,
\href{http://www.arXiv.org/abs/hep-ph/0507011}{{\tt hep-ph/0507011}}.

\bibitem{Moortgat-Picka:2015yla}
A.~Arbey {\em et al.}, {\em Eur. Phys. J.} {\bf C75} (2015), no.~8 371,
\href{http://www.arXiv.org/abs/1504.01726}{{\tt 1504.01726}}.

\bibitem{Badger:2005zh}
S.~D. Badger, E.~W.~N. Glover, V.~V. Khoze, and P.~Svrcek, {\em JHEP} {\bf 07}
  (2005) 025, \href{http://www.arXiv.org/abs/hep-th/0504159}{{\tt
  hep-th/0504159}}.

\bibitem{Badger:2005jv}
S.~D. Badger, E.~W.~N. Glover, and V.~V. Khoze, {\em JHEP} {\bf 01} (2006) 066,
  \href{http://www.arXiv.org/abs/hep-th/0507161}{{\tt hep-th/0507161}}.

\bibitem{MAITRE2008501}
D.~Ma\^{i}tre and P.~Mastrolia, {\em Computer Physics Communications} {\bf 179}
  (2008), no.~7 501 -- 534.

\bibitem{Bondarenko:2018sgg}
S.~Bondarenko, Y.~Dydyshka, L.~Kalinovskaya, L.~Rumyantsev, R.~Sadykov, and
  V.~Yermolchyk, {\em Phys. Rev. D} {\bf 100} (2019), no.~7 073002,
  \href{http://www.arXiv.org/abs/1812.10965}{{\tt 1812.10965}}.

\bibitem{Schwinn:2007ee}
C.~Schwinn and S.~Weinzierl, {\em JHEP} {\bf 04} (2007) 072,
  \href{http://www.arXiv.org/abs/hep-ph/0703021}{{\tt hep-ph/0703021}}.

\bibitem{Wigner:1957ep}
E.~P. Wigner, {\em Rev. Mod. Phys.} {\bf 29} (1957) 255--268.

\bibitem{Wigner:1964zf}
E.~P. Wigner, {\em Phys. Today} {\bf 17} (1964) 34--40.

\bibitem{Veltman:1977kh}
M.~Veltman, {\em Nuclear Physics B} {\bf 123} (1977), no.~1 89--99.

\bibitem{Fleischer:1993ub}
J.~Fleischer, O.~V. Tarasov, and F.~Jegerlehner, {\em Phys. Lett.} {\bf B319}
  (1993)
249--256.

\bibitem{Fleischer:1994cb}
J.~Fleischer, O.~V. Tarasov, and F.~Jegerlehner, {\em Phys. Rev.} {\bf D51}
  (1995)
3820--3837.

\bibitem{Bardin:280836}
D.~Y. Bardin, W.~Hollik, and G.~Passarino, {\em {Reports of the Working Group
  on precision calculations for the Z resonance}}.
\newblock CERN Yellow Reports: Monographs. CERN, Geneva, 1995.

\bibitem{Bardin:1999yd}
D.~{\relax Yu}. Bardin, P.~Christova, M.~Jack, L.~Kalinovskaya, A.~Olchevski,
  S.~Riemann, and T.~Riemann, {\em Comput. Phys. Commun.} {\bf 133} (2001)
  229--395,
\href{http://www.arXiv.org/abs/hep-ph/9908433}{{\tt hep-ph/9908433}}.

\bibitem{Barbieri:1992nz}
R.~Barbieri, M.~Beccaria, P.~Ciafaloni, G.~Curci, and A.~Vicere, {\em
  Phys.Lett.} {\bf B288} (1992) 95--98,
\href{http://www.arXiv.org/abs/hep-ph/9205238}{{\tt hep-ph/9205238}}.

\bibitem{Djouadi:1987gn}
A.~Djouadi and C.~Verzegnassi, {\em Phys. Lett.} {\bf B195} (1987)
265--271.

\bibitem{Djouadi:1987di}
A.~Djouadi, {\em Nuovo Cim.} {\bf A100} (1988)
357.

\bibitem{Consoli:1989fg}
M.~Consoli, W.~Hollik, and F.~Jegerlehner, {\em Phys. Lett. B} {\bf 227} (1989)
  167--170.

\bibitem{Consoli:1989pc}
M.~Consoli, W.~Hollik, and F.~Jegerlehner, ``{Electroweak Radiative Corrections
  for $Z$ Physics}'', in {\em {LEP Physics Workshop}}, 9, 1989.

\bibitem{Dittmaier:2009cr}
S.~Dittmaier and M.~Huber, {\em JHEP} {\bf 1001} (2010) 060,
\href{http://www.arXiv.org/abs/0911.2329}{{\tt 0911.2329}}.

\bibitem{Belyaev:2012qa}
A.~Belyaev, N.~D. Christensen, and A.~Pukhov, {\em Comput. Phys. Commun.} {\bf
  184} (2013) 1729--1769,
\href{http://www.arXiv.org/abs/1207.6082}{{\tt 1207.6082}}.

\bibitem{ParticleDataGroup:2020ssz}
{Particle Data Group} Collaboration, P.~A. Zyla {\em et al.}, {\em PTEP} {\bf
  2020} (2020), no.~8 083C01.

\end{thebibliography}\endgroup

\providecommand{\href}[2]{#2}

\end{document}